\documentclass[11pt]{article}

\usepackage{amsmath}
\usepackage{graphicx}
\usepackage{indentfirst}
\usepackage{amssymb}
\usepackage{cite}
\usepackage{color}
\usepackage{subfigure}
\usepackage{xcolor}
\usepackage[breaklinks=true]{hyperref}

\usepackage{mathrsfs}
\usepackage{amsmath}

\usepackage{orcidlink}

\begin{document}

\title{Shadows, Quasinormal Modes, and Optical Appearances of Black Holes in Horndeski Theory}

\date{}
\maketitle

\begin{center}
	\author{Zhi Luo\orcidlink{0009-0006-8622-2078}$~^{a,}$\footnote{xiyangdekehen11@gmail.com},
		Jin Li\orcidlink{0000-0001-8538-3714}$^{a,b,}$\footnote{cqstarv@hotmail.com},
		Ke-Jian He$~^{a,c,}$\footnote{kjhe94@163.com},
		Hao Yu\orcidlink{0000-0003-1689-2421}~$^{a,b,}$\footnote{yuhaocd@cqu.edu.cn (corresponding author)}}
\end{center}

\begin{center}
	$^a$ Department of Physics, Chongqing University, Chongqing 401331, China\\
	$^b$  Department of Physics and Chongqing Key Laboratory for Strongly Coupled Physics, Chongqing University, Chongqing 401331, China\\
	$^c$ Department of Mechanics, Chongqing Jiaotong University, Chongqing, 400074, China
\end{center}

\vskip 0.6in
{\abstract 
{
{This work describes the motion of photons in black hole (BH) spacetimes within the framework of Horndeski theory. We focus on the shadows, quasinormal modes (QNMs) and optical appearances of BHs surrounded by geometrically thin accretion disks. The QNMs of BHs are calculated by the WKB method and the eikonal limit, respectively. Using Event Horizon Telescope (EHT) observations of $\mathrm{M} 87^*$ and $\mathrm{Sgr} \mathrm{A}^*$, we can constrain the parameter in Horndeski theory to a small range. Based on the constraint, we obtain the frequency ranges of the fundamental modes for $\mathrm{M} 87^*$ and $\mathrm{Sgr} \mathrm{A}^*$ in Horndeski theory. By exploring the optical appearances of BHs, we find that for the current resolution of the EHT, it primarily captures direct emission. This work advances our understanding of the observational characteristics of BHs in Horndeski theory and constrains Horndeski theory by EHT observations of $\mathrm{M} 87^*$ and $\mathrm{Sgr} \mathrm{A}^*$.}
}


\thispagestyle{empty}
\newpage
\setcounter{page}{1}

\section{Introduction}\label{sec1}
In 2015, the Laser Interferometer Gravitational-Wave Observatory (LIGO) detectors made their first observation of gravitational waves (GWs), attributed to the merger of two stellar-mass black holes (BHs)~\cite{LIGOScientific:2016aoc,LIGOScientific:2016vlm}. This GW event provids evidence supporting the existence of BHs. Subsequent detections reveal more GW signals originating from the merger of different compact objects~\cite{LIGOScientific:2016sjg,LIGOScientific:2017vwq,LIGOScientific:2017bnn,LIGOScientific:2017ycc,LIGOScientific:2017vox}. These GW events have significantly broadened the scope of observational astronomy. In 2019, the Event Horizon Telescope (EHT) collaboration released the first image of the supermassive BH at the center of the $\mathrm{M} 87^*$ galaxy~\cite{EventHorizonTelescope:2019dse,EventHorizonTelescope:2019uob,EventHorizonTelescope:2019jan,EventHorizonTelescope:2019ths,EventHorizonTelescope:2019pgp,EventHorizonTelescope:2019ggy}. This image marked the first time in human history that a BH has been ``directly'' observed, which is currently  the most favorable evidence for the existence of BHs. Subsequently, the EHT collaboration published the image of the supermassive BH at the center of our Milky Way galaxy, $\mathrm{Sgr} \mathrm{A}^*$~\cite{EventHorizonTelescope:2022xnr,EventHorizonTelescope:2022vjs,EventHorizonTelescope:2022wok,EventHorizonTelescope:2022exc,EventHorizonTelescope:2022urf,EventHorizonTelescope:2022xqj}. The BH image depicts a dark region surrounded by a bright ring, which represents the BH shadow formed by the critical curve~\cite{Gralla:2019xty,Perlick:2021aok}. When photons from distant sources pass near a BH, their paths will bend due to the strong gravitational field around the BH. The photons with small impact parameters will fall into the BH, while those with large impact parameters can escape from the BH. For a given observer, the observed BH shadow mainly depends on two factors: the properties of the BH and the gravity theory. For example, a non-rotating BH could produce a circular shadow~\cite{Synge:1966okc,Luminet:1979nyg}, while a rotating BH exhibits a distorted shadow due to the dragging effect caused by its spin~\cite{Bardeen:1973tla,Hioki:2008zw,Eiroa:2017uuq}. In recent years, BH shadows in different modified gravity theories have been extensively studied~\cite{Amarilla:2010zq,Amarilla:2011fx,Amarilla:2013sj,Amir:2017slq,Singh:2017vfr,Mizuno:2018lxz,Vagnozzi:2019apd,Banerjee:2019nnj,Bambi:2008jg,Bambi:2010hf,Atamurotov:2013sca,Papnoi:2014aaa,Atamurotov:2015nra,Wang:2017qhh,Guo:2018kis,Yan:2019etp,Konoplya:2019sns,Bambi:2019tjh,Allahyari:2019jqz,Vagnozzi:2020quf,Khodadi:2020jij,Vagnozzi:2022moj,Tsukamoto:2014tja,Tsukamoto:2017fxq,Hu:2020usx,Zhong:2021mty,Peng:2020wun,Hou:2022eev,Hou:2021okc,Cunha:2018acu,Zeng:2020vsj,Zeng:2020dco,Zeng:2023zlf,Zeng:2021dlj,Olmo:2023lil,Asukula:2023akj,Saghafi:2022pme,Nozari:2023flq,EslamPanah:2024dfq,Jafarzade:2024knc,Hendi:2022qgi,EslamPanah:2020hoj}.

The final stage of GW signals, i.e., the ringdown phase, can be studied by BH perturbation theory, which is described by the quasinormal modes (QNMs). Generally, there are two approaches to investigate the QNMs of BHs: one is to consider a perturbation field, such as a massless scalar perturbation field, within the BH spacetime; and the other is to use a spectral decomposition of the metric perturbations. In the eikonal limit, the real part of the QNM frequencies is proportional to the angular velocity of the unstable circular null geodesic, while the imaginary part is linked to the Lyapunov exponent of the unstable circular null geodesic. Since the optical appearances of a BH are also related to the unstable circular null geodesic, the QNMs of the BH are naturally related to BH shadows~\cite{Ferrari:1984zz,Cardoso:2008bp,Hod:2009td,Jusufi:2019ltj,Stefanov:2010xz,Decanini:2002ha,Pedrotti:2024znu,Cai:2021ele}. Recently, the relationship between the real part of the QNM frequencies and the size of the BH
shadow has been studied in Refs.~\cite{Cuadros-Melgar:2020kqn,Jusufi:2020agr,Hendi:2020knv,Guo:2020nci,Jusufi:2020mmy,Jusufi:2020odz,Jusufi:2020wmp,Mondal:2020pop,Saurabh:2020zqg,Jafarzade:2020ilt,Jafarzade:2020ova,Ghasemi-Nodehi:2020oiz,Cai:2020igv,Campos:2021sff,Anacleto:2021qoe,Yu:2022yyv}. Furthermore, since real astrophysical BHs are always typically surrounded by significant amounts of accreting material, in order to study the optical appearances of real BHs, various accretion disk models have been proposed~\cite{Luminet:1979nyg,Gralla:2019xty,Gan:2021xdl,Bisnovatyi-Kogan:2022ujt,Gan:2021pwu,Cunha:2019hzj,Guo:2021bhr,Narayan:2019imo,Qin:2020xzu,Lambiase:2023hng}. Inspired by these works, we study the shadows, QNMs, and optical appearances of BHs surrounded by geometrically thin accretion disks in Horndeski theory~\cite{Horndeski:1974wa}. This work, on the one hand, helps us understand the observational characteristics of BHs in Horndeski theory, and on the other hand, can effectively constrain Horndeski theory through EHT observational data for  $\mathrm{M} 87^*$ and $\mathrm{Sgr} \mathrm{A}^*$.

Horndeski theory is the most general scalar-tensor theory characterized by second-order derivative field equations~\cite{Horndeski:1974wa}. Despite being proposed early on, Horndeski theory only regained attention when the covariant Galileon theory came up~\textit{et al.}~\cite{Nicolis:2008in,Deffayet:2009wt} (see Ref.~\cite{Kobayashi:2019hrl} for more details). Since then there have been some new developments in the gravity theory based on Horndeski theory. For example, in 2014, Gleyzes~\textit{et al.} proposed an extension to Horndeski theory, known as beyond Horndeski theory, aiming to mitigate Ostrogradski instabilities~\cite{Gleyzes:2014dya,Gleyzes:2014qga}. Subsequently, Babichev~\textit{et al.} discovered spherically symmetric and static BH solutions within both shift-symmetric Horndeski and beyond Horndeski theories~\cite{Babichev:2017guv}. In recent years, the studies of BHs in Horndeski theory and various beyond Horndeski theories have seen substantial scholarly interest~\cite{Rinaldi:2012vy,Cisterna:2014nua,Anabalon:2013oea,Babichev:2023psy,Charmousis:2014zaa,Minamitsuji:2013ura,Babichev:2016rlq,Babichev:2017guv,Feng:2015oea,Cvetic:2016bxi}. 

This paper is organized as follows. In Sec.~\ref{sec:Shadow}, we analyze the motion of photons in BH spacetimes within the framework of Horndeski theory, deriving the effective potential, critical impact parameter, and photon sphere radius. We explore the influences of the parameter in Horndeski theory on these variables and use EHT observational data for $\mathrm{M} 87^*$ and $\mathrm{Sgr} \mathrm{A}^*$ to constrain the parameter in Horndeski theory. Then, in Sec.~\ref{sec:4}, we study the perturbations of a massless scalar field in BH spacetimes. The QNMs of BHs are calculated by the WKB method and the eikonal limit, respectively. We study the relationship between the QNMs and the BH shadow radius. In Sec.~\ref{sec:QNMs}, we study the optical appearances of the BH surrounded by a geometrically thin accretion disk and compare the numerical results with the images of $\mathrm{M} 87^*$ and $\mathrm{Sgr} \mathrm{A}^*$ given by the EHT. At Last, Sec.~\ref{sec:5} is the conclusions of the work. Throughout this paper, we adopt natural units $8\pi G = 1 = c  $.

\section{BH shadows in Horndeski theory}\label{sec:Shadow}
We start from the Horndeski action~\cite{Horndeski:1974wa}
\begin{equation}\label{eq1}
	S=\int d^4 x \sqrt{-g}\left(\mathcal{L}_2+\mathcal{L}_3+\mathcal{L}_4+\mathcal{L}_5\right),
\end{equation}
where
\begin{equation}
	\mathcal{L}_2=G_2(X), \quad \mathcal{L}_3=-G_3(X) \square \phi,
\end{equation}
\begin{equation}
	\mathcal{L}_4=G_4(X) \mathcal{R}+G_{4 X}\left[(\square \phi)^2-\left(\nabla_\mu \nabla_\nu \phi\right)^2\right],
\end{equation}
\begin{equation}
	\mathcal{L}_5=G_5(X) G_{\mu \nu} \nabla^\mu \nabla^\nu \phi-\frac{G_{5 X}}{6}\left[(\square \phi)^3-3 \square \phi\left(\nabla_\mu \nabla_\nu \phi\right)^2+2\left(\nabla_\mu \nabla_\nu \phi\right)^3\right].
\end{equation}
Within this framework, $\mathcal{R}$ represents the Ricci scalar, $G_{\mu \nu}$ is the Einstein tensor, and $G_i=G_i(\phi, X)$ is the function of the scalar field $\phi$ and its kinetic term $X=-\frac12\partial_\mu \phi \partial^\mu \phi $. The subscript $X$ denotes the derivative with respect to $X$.
Following Ref~\cite{Babichev:2017guv}, we consider a special case of the Horndeski action (\ref{eq1}), which is described as
\begin{equation}
	G_2 = \eta X, \quad G_4 = \zeta + \beta \sqrt{-X}, \quad G_3 = G_5 = 0. 
\end{equation}
Here, we have $\zeta=\frac1{2}$ in natural units ($8\pi G = 1 = c $). Note that $\eta$ and $\beta$ are dimensionless parameters, whose values currently can only be determined through observations or experiments. Therefore, the action (\ref{eq1}) is reduced to
\begin{equation}
	\begin{aligned}
		S= & \int \mathrm{d}^4 x \sqrt{-g}\left\{\left[\frac1{2}+\beta \sqrt{(\partial \phi)^2 / 2}\right] \mathcal{R}-\frac{\eta}{2}(\partial \phi)^2-\frac{\beta}{\sqrt{2(\partial \phi)^2}}\left[(\square \phi)^2-\left(\nabla_\mu \nabla_\nu \phi\right)^2\right]\right\}.
	\end{aligned}
\end{equation}
For the spherically symmetric solution of the BH  in Horndeski theory, the corresponding line element takes the form~\cite{Babichev:2017guv}     
\begin{equation}\label{eqdsff}
	\mathrm{d} s^2=-f(r) \mathrm{d} t^2+\frac{1}{f(r)} \mathrm{d} r^2+r^2\left(\mathrm{d} \theta^2+\sin ^2 \theta \mathrm{d} \phi^2\right),
\end{equation}
where
\begin{equation}\label{eq5ff}
	f(r)=1-\frac{2M}{r}-\frac{\beta^2}{ \eta\, r^2}.
\end{equation}
Here, $M$ is an integration constant that represents the BH mass. Since both parameters $\eta$ and $\beta$ are dimensionless and independent, one can regard $\beta^2 /\eta$ as a whole. In this work, we define a new parameter $\gamma=\beta^2 /\eta$ to simplify the following discussion. Then, we can reformulate Eq.~(\ref{eq5ff}) as
\begin{equation}\label{eqfgamma}
	f(r)=1-\frac{2M}{r}-\frac{\gamma}{r^2}.
\end{equation}
When the parameter $\gamma$ takes a negative value, $\sqrt{-\gamma}$ behaves similarly to the electric charge of the Reissner-Nordström BH. The radius of the outer event horizon can be determined by solving the equation $f(r_h) = 0$, yielding $r_h = M + \sqrt{M^2 + \gamma}$. To avoid the naked singularity, we need $\gamma \geq -M^2$. For $\gamma = -M^2$, the solution~(\ref{eqdsff}) corresponds to an extremal BH. The metric function~(\ref{eqfgamma}) is illustrated in Fig.~\ref{frr}. It is found that for $\gamma > 0$ and $\gamma = 0$ (the Schwarzschild case), there is only one event horizon. However, for $-M^2 < \gamma < 0$, both inner and outer horizons are present.

\begin{figure}[!ht]
	\centering
	{
		\begin{minipage}[t]{0.5\linewidth}
			\centering
			\includegraphics[width=6cm,height=6cm]{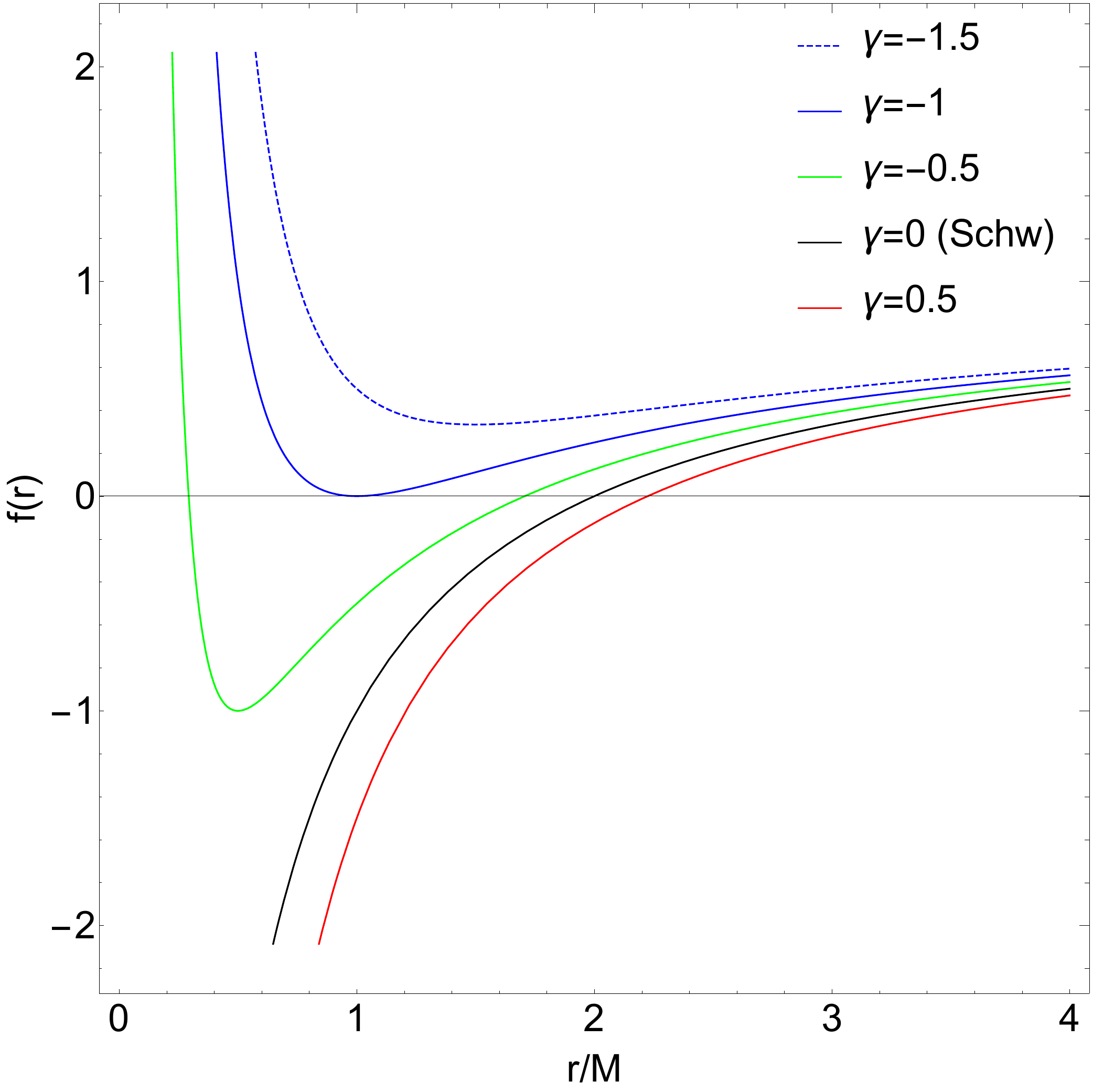}
		\end{minipage}%
	}%
	\caption{The metric function $f(r)$ for different values of the parameter $\gamma$. We set $ \gamma$= -1.5, -1, -0.5, 0, 0.5 and $M=1$.} 
	\label{frr}
\end{figure}

To study the BH shadow in Horndeski theory, we use the Lagrangian of photons to derive the equation of motion of photons, which is given by
\begin{equation}
	\label{Lagrangian}
	\mathcal{L}= \frac{1}{2} g_{\mu \nu} \dot{x}^\mu \dot{x}^\nu = \frac{1}{2}\left(-f(r) \dot{t}^2+\frac{\dot{r}^2}{f(r)}+r^2\left(\dot{\theta}^2+\sin ^2 \theta \dot{\phi}^2\right)\right).
\end{equation}
{Here, $\dot{x}^{\mu}\equiv {\rm d} x^{\mu}/{\rm d} \lambda$ is the four-velocity} of photons and $\lambda$ is the affine parameter. Considering a photon moving on the equatorial plane (i.e. $\theta=\pi/2$ and $\dot{\theta}=0$), since the the metric function $f(r)$ does not depend on the time coordinate $t$ and the azimuthal angle $\phi$, there are two conserved quantities:
\begin{equation}
	\label{conserved}
	p_{t}=\frac{\partial \mathcal{L} }{\partial \dot{t}}= f(r) \dot{t}= E, \quad p_{\phi}=-\frac{\partial \mathcal{L} }{\partial \dot{\phi}}= r^{2} \dot{\phi} = L.
\end{equation}
In addition, we can obtain 
\begin{eqnarray}
	\label{fourv}
	\dot{t}=\frac{1}{b}f(r)^{-1}, \quad \dot{r}=\sqrt{\frac{1}{b^{2}} - \frac{1}{r^{2} }f(r)}, \quad \dot{\phi}=\pm\frac{1}{r^{2}},
\end{eqnarray}
where ``$\pm$'' indicates the direction of the photon's motion (``$+$'' represents clockwise and ``$-$'' represents counterclockwise). The parameter $b = |L|/E$ is called the impact parameter. Using Eq.~(\ref{fourv}), we can obtain    
\begin{equation}
	\dot{r}^2+V_{\text{eff}}=\frac{1}{b^2},
\end{equation}
where $V_{\text{eff}}=\frac{f(r)}{r^2} $ denotes the effective potential of the photon. The critical conditions for the photon's unstable circular orbit are
\begin{equation}\label{vvp}
	V_{\text{eff}}\left(r_{\text{ph}}\right)=\frac{1}{b_c^2}, \quad V_{\text{eff}}^{\prime}(r_{\text{ph}})=0,
\end{equation}
where $r_{\text{ph}}$ is the radius of the photon sphere and $b_c$ is the critical impact parameter. With the solution~(\ref{eqdsff}), we can obatin the analytical expressions of $r_{\text{ph}}$ and $b_c$:
\begin{equation}
	r_{\text{ph}}=\frac{1}{2}\left(3 M+\sqrt{9 M^2+8 \gamma}\right),
\end{equation}
\begin{equation}\label{bcc}
	b_c=\frac{\sqrt{27 M^4+36 M^2 \gamma+8 \gamma^2+9 M^3 \sqrt{9 M^2+8 \gamma}+8 M \gamma \sqrt{9 M^2+8 \gamma}}}{\sqrt{2} \sqrt{M^2+\gamma}}.
\end{equation} 
For a given $M$, the photon sphere radius $r_{\text{ph}}$ and the critical impact parameter $b_c$ only depend on the parameter $\gamma$. From Fig.~\ref{bcrprh}, one can find that the event horizon radius, photon sphere radius, and critical impact parameter all increase with the parameter $\gamma$. We plot the effective potential of the photon for different values of the parameter $\gamma$ in Fig.~\ref{veff}. It is found that the effective potential (e.g. $\gamma = -0.2$) rises from the event horizon radius, reaches a peak at the photon sphere radius, and then declines. Additionally, as the parameter $\gamma$ increases, the peak value of the effective potential will decrease.

\begin{figure}[!ht]
	\centering
		\begin{minipage}[t]{0.5\linewidth}
			\centering
			\includegraphics[width=6cm,height=6cm]{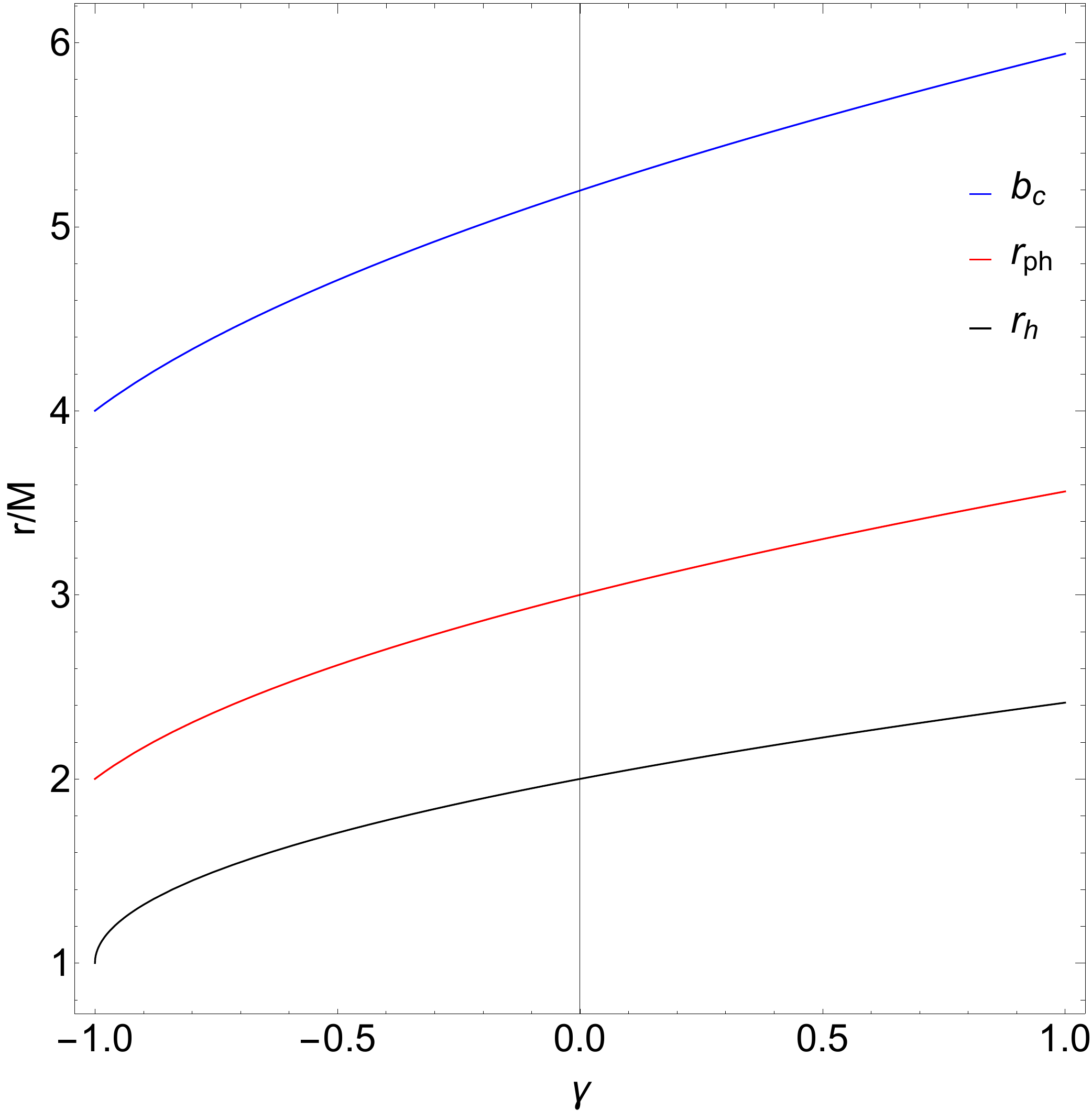}
		\end{minipage}%
	\caption{The event horizon radius, photon sphere radius, and critical impact parameter vary with the parameter $\gamma$. We set $M=1$.} 
	\label{bcrprh}
\end{figure}

\begin{figure}[!ht]
	\centering
		\begin{minipage}[t]{0.5\linewidth}
			\centering
			\includegraphics[width=6cm,height=6cm]{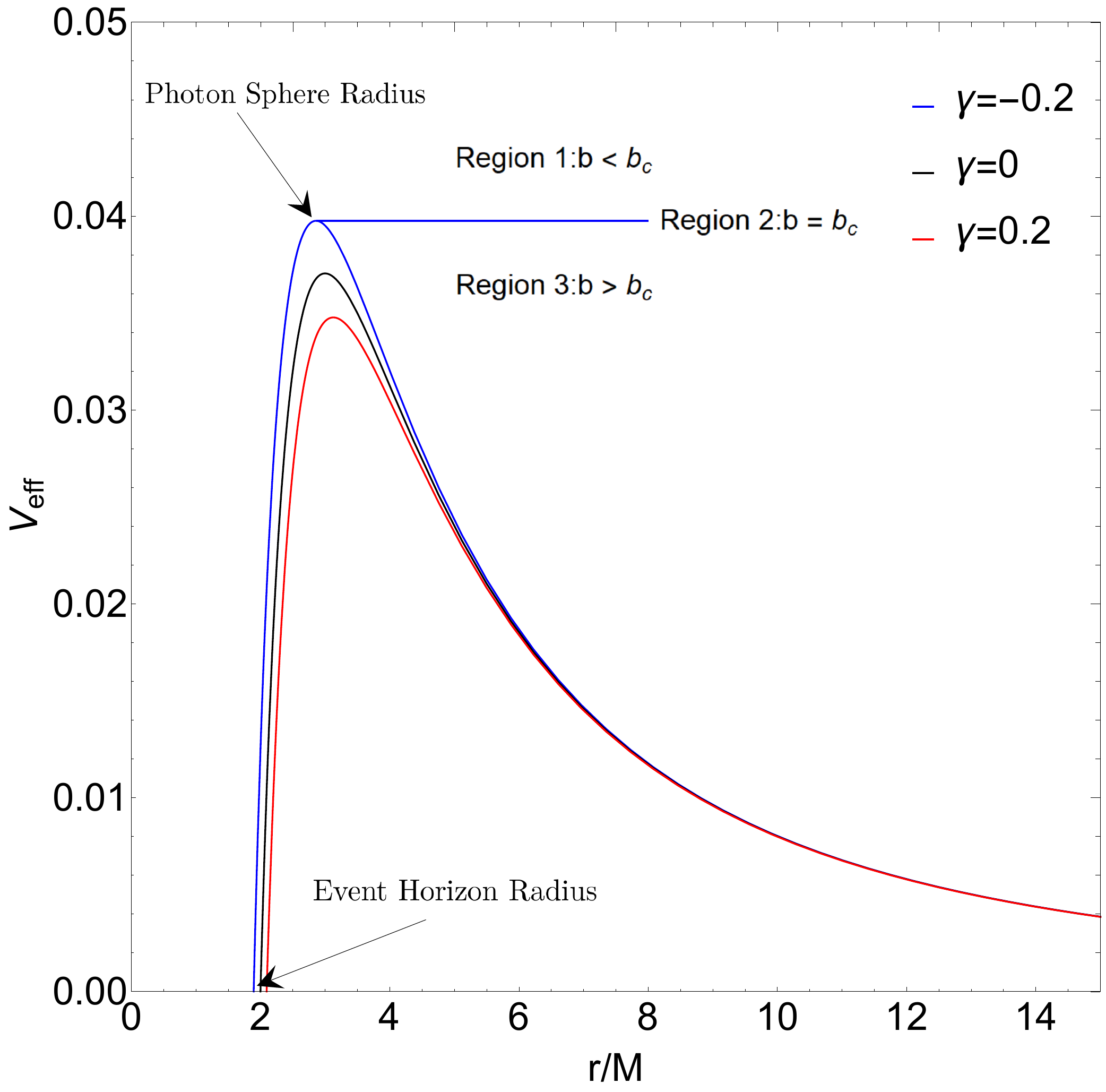}
		\end{minipage}%
	\caption{The effective potential of the photon for different values of the parameters $\gamma$. We set $ \gamma$= -0.2, 0, 0.2 and $M=1$.} 
	\label{veff}
\end{figure}

To study the trajectory of the photon, we need to rewrite the photon's equation of motion using Eq.~(\ref{fourv}), which can be written as 
\begin{equation}\label{drdphi}
	\frac{d r}{d \phi}= \pm r^2 \sqrt{\frac{1}{b^2}-\frac{1}{r^2}\left(1-\frac{2 M}{r}-\frac{\gamma}{r^2}\right)}.
\end{equation}
Through a coordinate transformation $x \equiv 1/r$, Eq.~(\ref{drdphi}) can be further  expressed as
\begin{equation}\label{drdphinew}
	\mathcal{G}(x, b) \equiv \frac{d x}{d \phi}=\sqrt{\frac{1}{b^2}-x^2\left(1-2 M x-\gamma x^2\right)}.
\end{equation}
When the photon with an impact parameter $b$ approaches the BH from infinity, there are three possible scenarios based on the value of $b$~\cite{Gralla:2019xty}:
\begin{itemize}
	\item \emph{$b < b_c$}~~ The photon will finally fall into the BH. If $b=0$, the trajectory of the photon is a straight line. If $0<b<b_c$, there is a deflection angle for the trajectory of the photon. For the event horizon $r_{h}  $, the total deflection angle of the photon is given by
	\begin{equation}\label{eq19y}
		\phi=\int_0^{1/r_h} \frac{\mathrm{d} x}{\sqrt{\mathcal{G}(x, b)}} .
	\end{equation}
	\item \emph{$b = b_c$}~~ The photon will reach a critical point $x_c$ and then orbit the BH along a circular path with a radius of $r_{\text{ph}}$. In this scenario, the motion of the photon is in a critical state, which delineates the boundary of the BH shadow.
	\item \emph{$b > b_c$}~~ The photon will reach a nearest point $x_{\text{min}}$ (the smallest positive real root of $\mathcal{G}(x, b) = 0$) and then escape to infinity. The corresponding total deflection angle is given by
	\begin{equation}
	\phi=2 \int_0^{x_{\min }} \frac{\mathrm{d} x}{\sqrt{\mathcal{G}(x, b)}} .
	\end{equation}
\end{itemize}
In Fig.~\ref{traj}, we plot the trajectories of photons with different impact parameters. Here, the impact parameter $b$ can be considered as the vertical distance between the photon's geodesic and the parallel line intersecting the origin. When $b<b_c$ (see Region 1 in Fig.~\ref{veff} and the black lines in Fig.~\ref{traj}), photons will be captured by the BH. When $b=b_c$ (see Region 2 in Fig.~\ref{veff} and the red rings in Fig.~\ref{traj}), photons will continue to orbit the BH. When $b>b_c$ (see Region 3 in Fig.~\ref{veff} and the blue lines in Fig.~\ref{traj}), photons will be reflected back and move to infinity.
\begin{figure}[!ht]
	\centering
	\subfigure[$\gamma = -0.3$\label{traj1}]{
		\begin{minipage}[t]{0.33\linewidth}
			\centering
			\includegraphics[width=4.5cm,height=4.5cm]{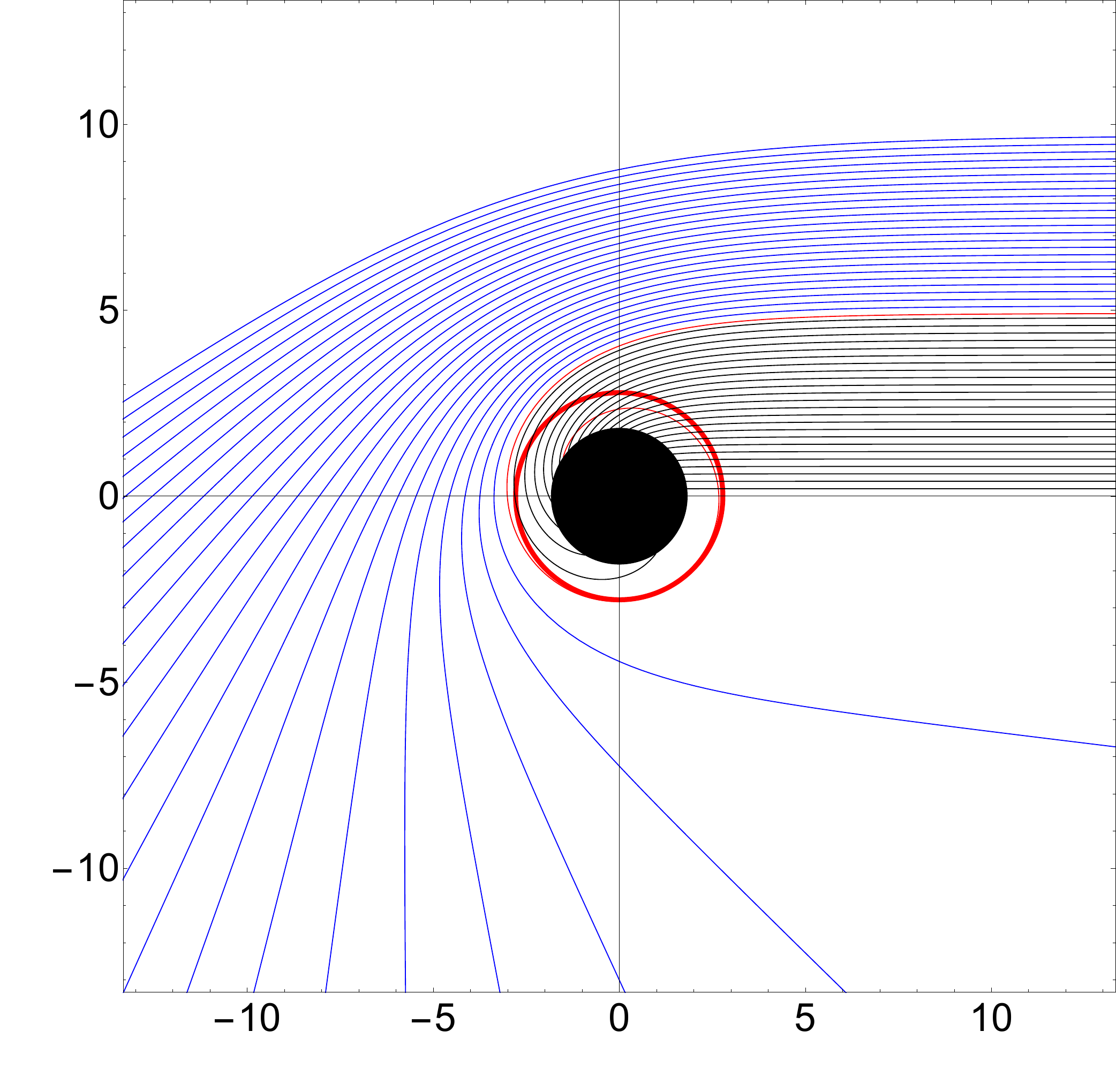}
		\end{minipage}%
	}%
	\subfigure[$\gamma = 0$\label{traj2}]{
		\begin{minipage}[t]{0.33\linewidth}
			\centering
			\includegraphics[width=4.5cm,height=4.5cm]{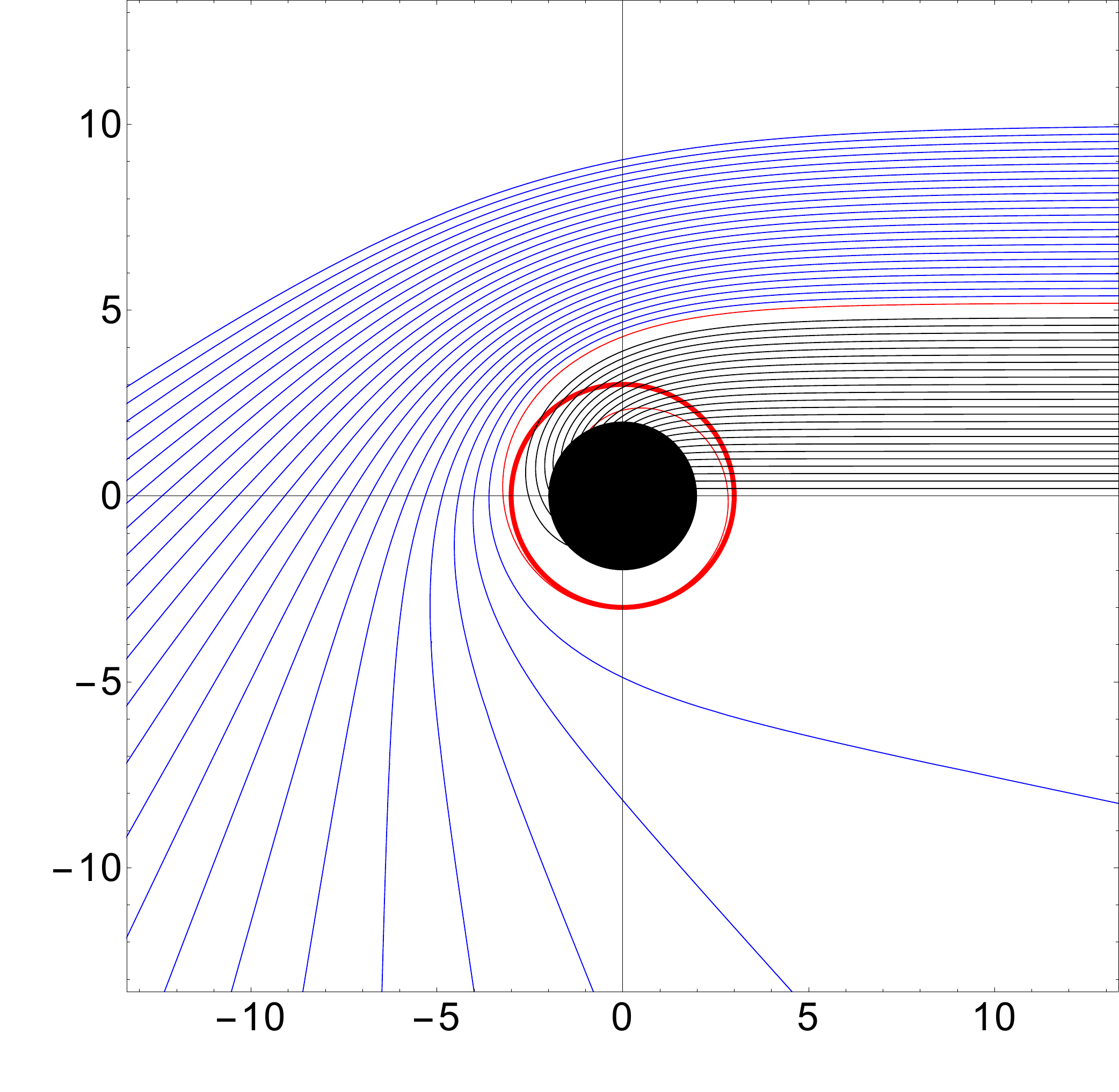}
		\end{minipage}%
	}%
	\subfigure[$\gamma = 0.3$\label{traj3}]{
		\begin{minipage}[t]{0.33\linewidth}
			\centering
			\includegraphics[width=4.5cm,height=4.5cm]{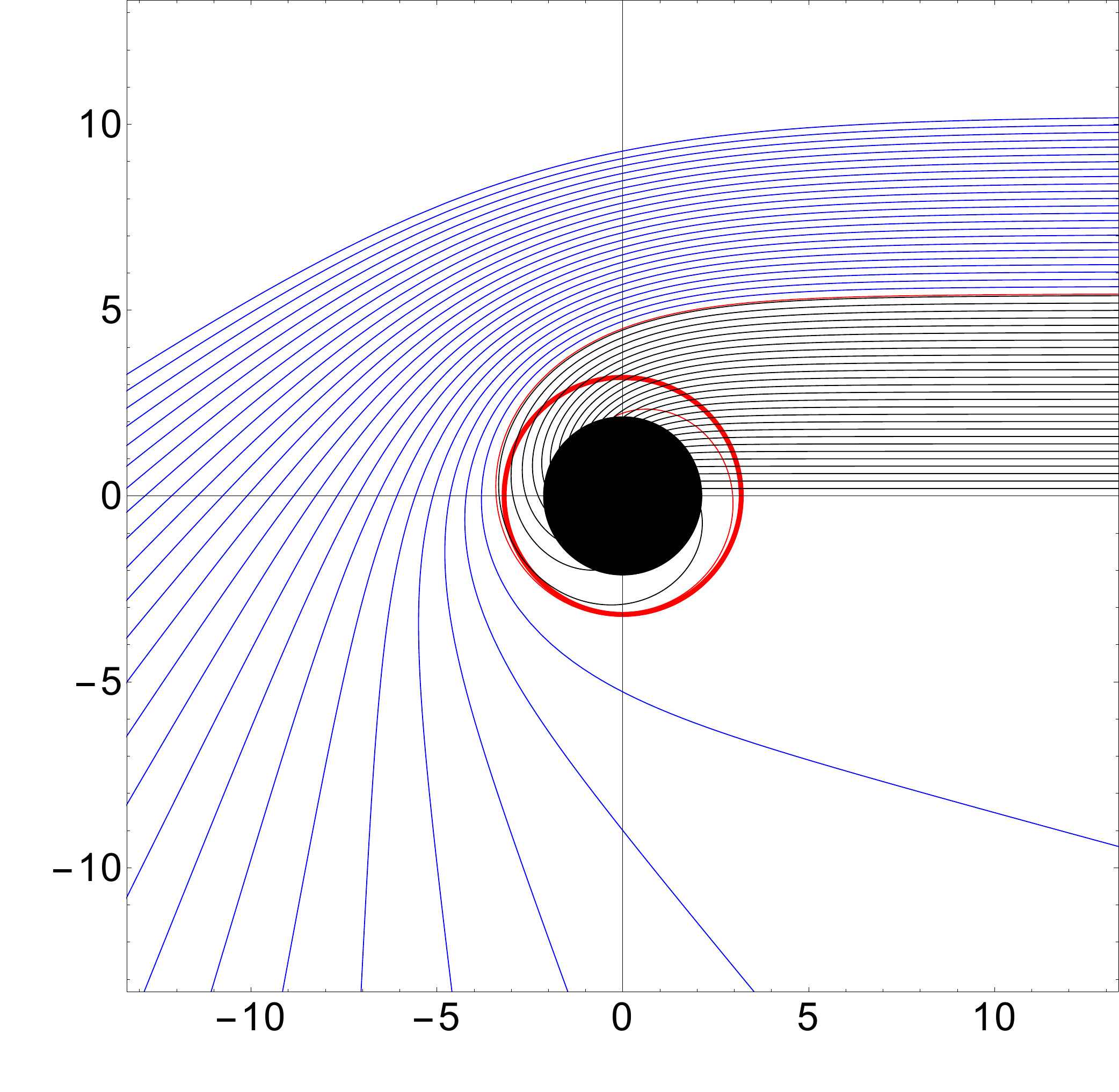}
		\end{minipage}
	}%
	\caption{The trajectories of photons moving on the equatorial plane of the BH for different values of the parameter $\gamma$ in polar coordinates $(r, \phi)$. The BH is depicted as a black disk in the center. The red rings represent the critical curve. The black lines and blue lines correspond to $b < b_c$ and $b > b_c$, respectively.  We set $\gamma$ = -0.3, 0, 0.3 and $M = 1$. The spacing of the impact parameter is set to $1/5$. }
	\label{traj}
\end{figure}

With the calculations above, we can constrain the parameter $\gamma$ using the observed shadows of the supermassive BH $\mathrm{M} 87^*$ and $\mathrm{Sgr} \mathrm{A}^*$ captured by the EHT collaboration~\cite{EventHorizonTelescope:2019dse,EventHorizonTelescope:2019uob,EventHorizonTelescope:2019jan,EventHorizonTelescope:2019ths,EventHorizonTelescope:2019pgp,EventHorizonTelescope:2019ggy,EventHorizonTelescope:2022xnr,EventHorizonTelescope:2022vjs,EventHorizonTelescope:2022wok,EventHorizonTelescope:2022exc,EventHorizonTelescope:2022urf,EventHorizonTelescope:2022xqj}. For a static observer located at position $r_0$, the radius of the BH shadow can be expressed as~\cite{EventHorizonTelescope:2022xqj,Perlick:2015vta,EventHorizonTelescope:2020qrl} 
\begin{equation}\label{Rsh}
	\mathcal{R}_{\text {sh }}=r_{\mathrm{ph}} \sqrt{\frac{f\left(r_0\right)}{f\left(r_{\mathrm{ph}}\right)}},
\end{equation}
where $f\left(r_0\rightarrow\infty\right)=1$. In units of the BH mass, the diameter of the BH shadow in the practical observetions is given by
\begin{equation}
	d_{\mathrm{sh}}=2\mathcal{R}_{\text {sh }}=\frac{D \theta} {M},
\end{equation}
where $\theta$ is the angular diameter of the BH shadow and $D$ is the distance between  the BH and the observer at position $r_0$. According to Refs.~\cite{EventHorizonTelescope:2019dse,EventHorizonTelescope:2022wok}, the angular diameter of the shadow for $\mathrm{M} 87^*$ is $\theta_{\mathrm{M} 87^*} = (42\pm3)$ $\mu$as. The distance from the Earth is $D_{\mathrm{M} 87^*}= 16.8_{-0.7}^{+0.8}$ Mpc and the mass is $M_{\mathrm{M} 87^*}=(6.5 \pm 0.9) \times 10^{9}M_{\odot}$. For $\mathrm{Sgr} \mathrm{A}^*$, the angular diameter is $\theta_{\text {SgrA*}}=(48.7\pm7)$ $\mu$as. The distance is $D_{\text {SgrA*}}= 8277 \pm 33$ {pc} and the mass is $M_{\text {SgrA*}}=(4.3 \pm 0.013) \times 10^6 M_{\odot}$. Hence, the diameters of the shadow images of $\mathrm{M} 87^*$ and $\mathrm{Sgr} \mathrm{A}^*$ are $d_{\mathrm{sh}}^{\mathrm{M} 87^*}=(11 \pm 1.5)$ and $d_{\mathrm{sh}}^{\text {SgrA*}}=(9.5 \pm 1.4)$, respectively. Taking these results into Eq.~(\ref{Rsh}), then we can constrain the parameter $\lambda$ (see Fig.~\ref{constrain} for details). Based on the confidence levels for $d_{\mathrm{sh}}^{\mathrm{M} 87^*}$, the parameter $\gamma$ is constrained to $-0.46 \lesssim \gamma \lesssim 1.50$ at the $1\sigma$ confidence level and $-1.00 \lesssim \gamma \lesssim 2.92$ at the $2\sigma$ confidence level. For $d_{\mathrm{sh}}^{\text {SgrA*}}$, the parameter $\gamma$ is constrained to $-0.97 \lesssim \gamma \lesssim 0.31$ at the $1\sigma$ confidence level and $-1.12 \lesssim \gamma \lesssim 1.34$ at the $2\sigma$ confidence level. Apparently, the constraint on the upper limit of the parameter $\gamma$ obtained from $\mathrm{Sgr} \mathrm{A}^*$ is stricter than that obtained from $\mathrm{M} 87^*$. But the constraint on the lower limit of the parameter $\gamma$ from $\mathrm{M} 87^*$ is more stringent than that from $\mathrm{Sgr} \mathrm{A}^*$.
\begin{figure}[!ht]
	\centering
	\subfigure[Constraining the parameter $\gamma$ with $d_{\mathrm{sh}}^{\mathrm{M} 87^*}$ \label{constrain1}]{
		\begin{minipage}[t]{0.5\linewidth}
			\centering
			\includegraphics[width=6cm,height=6cm]{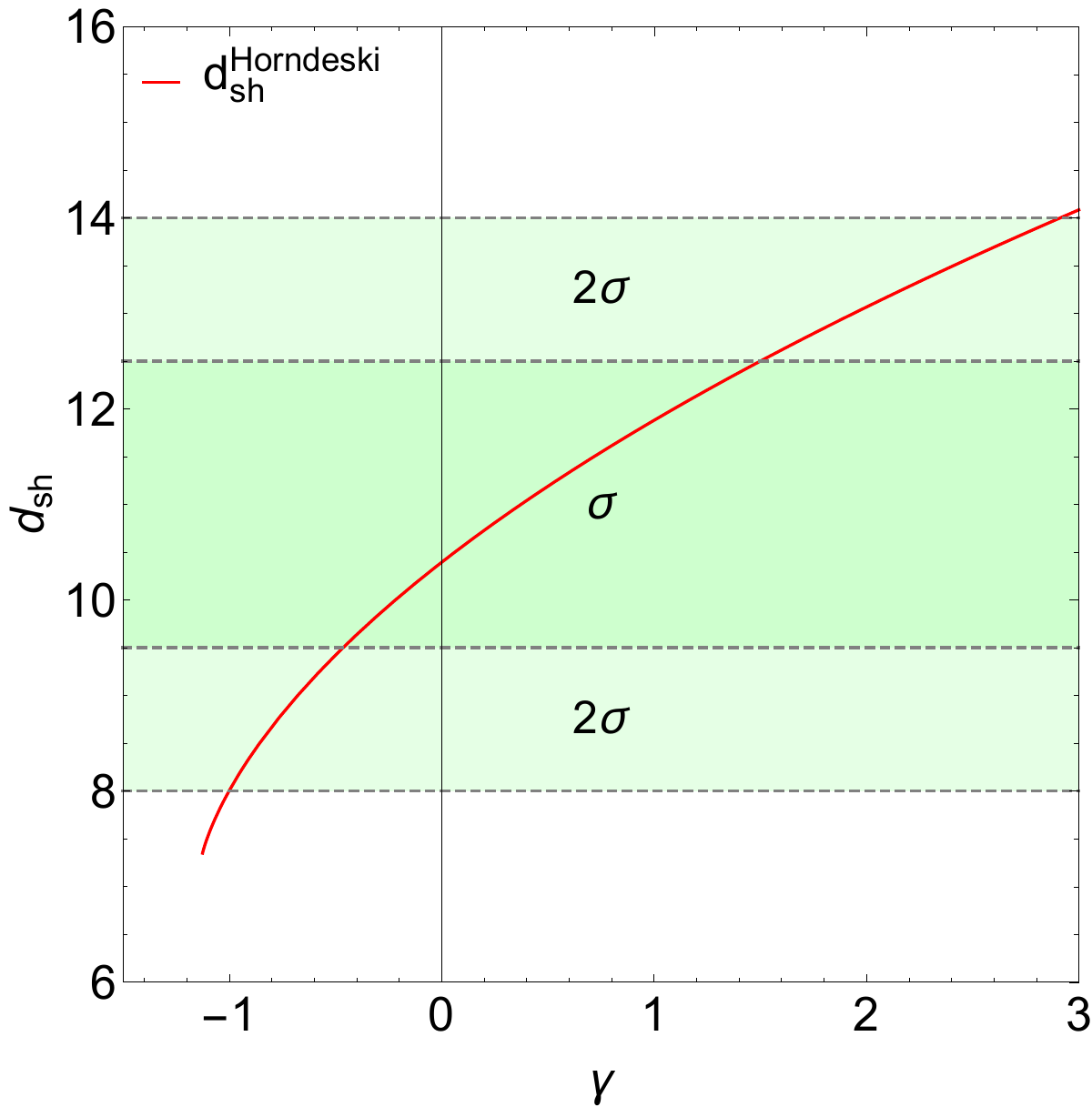}
		\end{minipage}%
	}%
	\subfigure[Constraining the parameter $\gamma$ with $d_{\mathrm{sh}}^{\text {SgrA*}}$ \label{constrain2}]{
		\begin{minipage}[t]{0.5\linewidth}
			\centering
			\includegraphics[width=6cm,height=6cm]{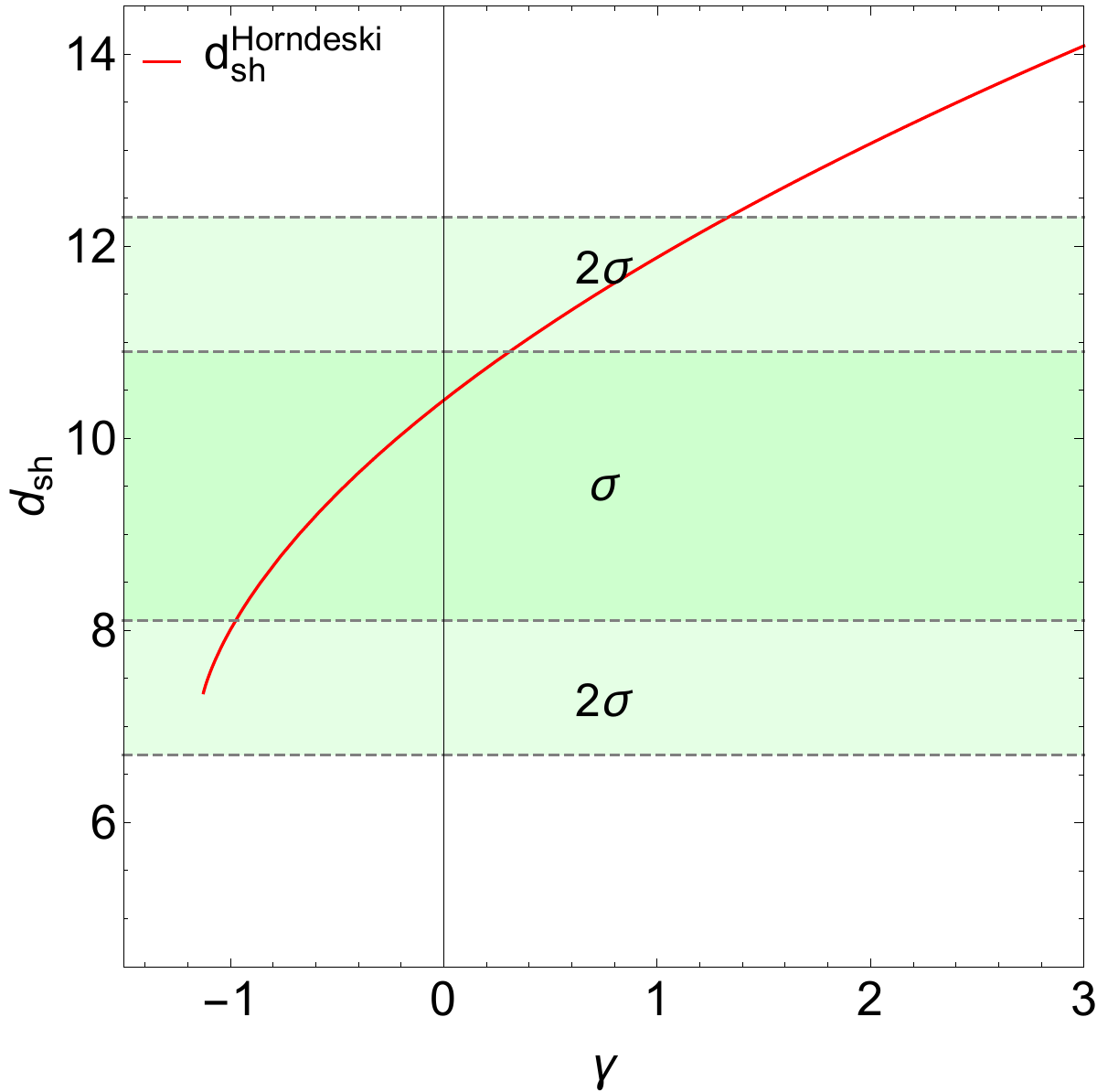}
		\end{minipage}%
	}%
	\caption{The diameter $d_{\mathrm{sh}}^{\text {Horndeski}}$ (the red lines) of the BH shadow varies with the parameter $\gamma$. The shaded areas represent the value of $d_{\mathrm{sh}}^{\text {Horndeski}}$ that are consistent with the observed shadow diameters of $\mathrm{M} 87^*$ (the left panel) and $\mathrm{Sgr}  \mathrm{A}^*$ (the right panel) given by the EHT. The green shaded areas indicate the $1\sigma$ confidence level for $d_{\mathrm{sh}}^{\text {Horndeski}}$, while the light green shaded areas indicate the $2\sigma$ confidence level. We set $M = 1$.} 
	\label{constrain}
\end{figure}\textbf{}

\section{QNMs of a massless scalar perturbation field}\label{sec:QNMs}
In this section, we study the QNMs of a massless scalar perturbation field in the geometry of the BH in Horndeski theory. We employ the 6th-order WKB method and the eikonal limit to calculate the QNMs, respectively. We use the EHT data presented in Sec.~\ref{sec:Shadow} to determine the oscillation frequencies of a massless scalar perturbation field for $\mathrm{M} 87^*$ and $\mathrm{Sgr} \mathrm{A}^*$.

\subsection{QNMs with the 6th-order WKB method}\label{sec:QNMs1}
The Klein-Gordon equation of a massless scalar perturbation field $\Phi$ in the geometry of the BH in Horndeski theory is given by
\begin{equation}\label{massless}
	\frac{1}{\sqrt{-g}} \partial_\mu\left(\sqrt{-g} g^{\mu \nu} \partial_\nu \Phi\right)=0
\end{equation}
with
\begin{equation}
	\Phi=e^{-i \omega t} Y_{\ell m}(\theta, \varphi) \frac{\Psi(r)}{r},
\end{equation}
where $g_{\mu \nu }$ is the metric tensor (\ref{eqdsff}). Through a coordinate transformation $dr_{*} \equiv dr/f(r)$, we can get
\begin{equation}
	\frac{\mathrm{d}^2 \Psi\left(r_*\right)}{\mathrm{d} r_*^2}+\left[\omega^2-V_S(r)\right] \Psi\left(r_*\right)=0
\end{equation}
with
\begin{equation}
	V_S(r)=f(r)\left(\frac{\ell(\ell+1)}{r^2}+\frac{f^{\prime}(r)}{r} \right).
\end{equation}
Here, $V_S(r)$ is the effective potential of the massless scalar perturbation field, $\ell$ is the multipole number, and $\omega$ is the complex quasinormal mode frequency. We use the $6$th-order WKB method to calculate the QNM frequencies of the massless scalar perturbation field. The complex quasinormal mode frequency $\omega$ satisfies~\cite{Schutz:1985km,Iyer:1986np,Konoplya:2003ii}
\begin{equation}
	\frac{\mathrm{i}\left(\omega^2-V_{max}\right)}{\sqrt{-2 V_{max}^{\prime \prime}}}-\sum_{i=2}^6 \Lambda_i=\mathfrak{n}+\frac{1}{2},
\end{equation}
where $V_{max}$ and $V_{max}^{\prime \prime}$ represent the maximum values of the effective potential and its second derivative with respect to $r_*$, respectively. The parameter $\mathfrak{n}$ is the overtone number and $\Lambda_i$ is the $i$th-order correction term given by Refs.~\cite{Schutz:1985km,Iyer:1986np,Konoplya:2003ii}. In the following study, we focus on the fundamental QNMs with the overtone number $\mathfrak{n}=0$ and the multipole number $\ell=2$. Moreover, based on our previous calculations of $\mathrm{M} 87^*$ and $\mathrm{Sgr} \mathrm{A}^*$, and considering the $1\sigma$ confidence level, the parameter $\gamma$ can be strictly constrained within the range $-0.46 \lesssim \gamma \lesssim 0.31$. Therefore, to simplify the problem, our subsequent discussions regarding $\gamma$ will be confined to this range.

The real and negative imaginary parts of the QNM frequencies are shown in Fig.~\ref{qnms1}. It is found that as the parameter $\gamma$ increases, both the real part and the negative imaginary part decrease monotonically. This implies that for smaller values of the parameter $\gamma$, the scalar wave undergoes more frequent oscillations.

\begin{figure}[!ht]
	\centering
	\subfigure[The real parts of the QNM frequencies \label{qnm1}]{
		\begin{minipage}[t]{0.5\linewidth}
			\centering
			\includegraphics[width=6cm,height=6cm]{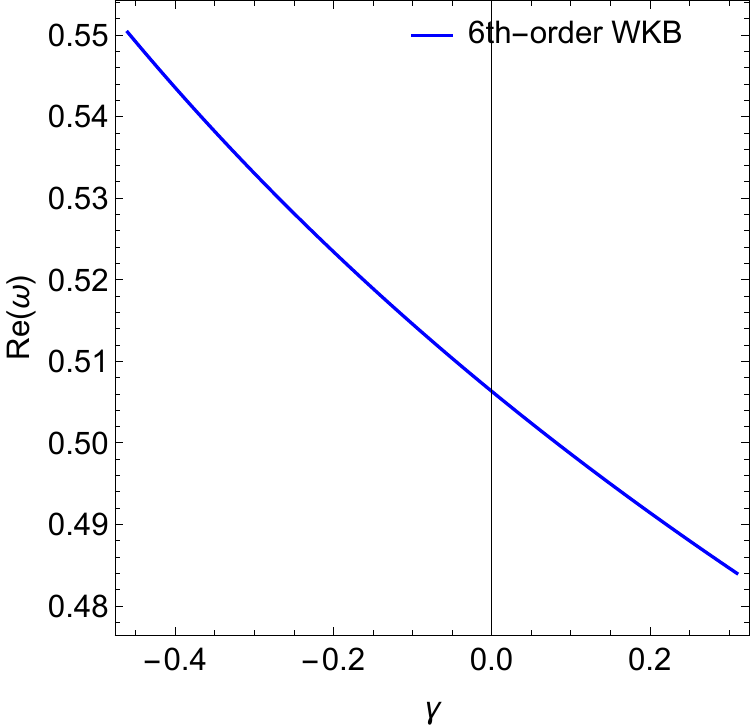}
		\end{minipage}%
	}%
	\subfigure[The negative imaginary parts of the QNM frequencies \label{qnm3}]{
		\begin{minipage}[t]{0.5\linewidth}
			\centering
			\includegraphics[width=6cm,height=6cm]{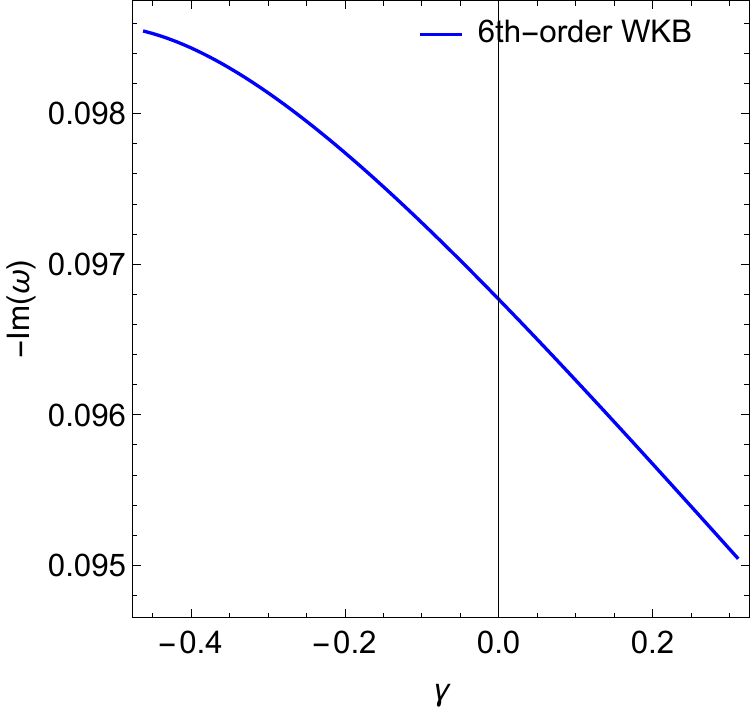}
		\end{minipage}%
	}%
	\caption{The real and negative imaginary parts of the QNM frequencies calculated via the $6$th-order WKB method. We set the overtone number $\mathfrak{n}=0$, the multipole number $\ell=2$, and $M = 1$.} 
	\label{qnms1}
\end{figure}

\subsection{QNMs in the eikonal limit}\label{sec:QNMs2}
In the eikonal limit $\ell \gg 1$, Cardoso et al.~\cite{Cardoso:2008bp} suggest another way of calculating the QNMs of BHs with unstable circular null geodesics. In this method, the complex quasinormal mode frequency $\omega$ is given by~\cite{Cardoso:2008bp,Konoplya:2017lhs,Konoplya:2017wot}  
\begin{equation}
	\omega=\Omega_c \ell-\mathrm{i}(\mathfrak{n}+1 / 2)\left|\lambda_{\mathrm{L}}\right|,
\end{equation}
where
\begin{equation}
	\label{Omegac}
	\Omega_c=\frac{\sqrt{f(r_c)}}{r_c},        \qquad   \lambda_{\mathrm{L}}=\sqrt{\frac{f\left(r_c\right)\left[2 f\left(r_c\right)-r_c^2 f^{\prime \prime}\left(r_c\right)\right]}{2 r_c^2}}.
\end{equation}
Here, $\Omega_c$ is the angular velocity and $\lambda_{\mathrm{L}}$ is the Lyapunov exponent of the unstable circular null geodesics. The radius of the circular null geodesic, denoted as $r_c$, satisfies the condition $2 f(r_c)=r_c f(r_c)^{\prime}$. In Fig.~\ref{qnms2}, we plot the real and negative imaginary parts of the QNM frequencies via the eikonal limit. As the parameter $\gamma$ increases, both the real part and the negative imaginary part of the QNM frequencies monotonically decrease, which is consistent with the results shown in Fig.~\ref{qnms1} via the $6$th-order WKB method. 

The relationship between the angular velocity $\Omega_c$ in Eq.~(\ref{Omegac}) and the shadow radius $\mathcal{R}_{\text {sh }}$ in Eq.~(\ref{Rsh}) is given by~\cite{Lu:2019zxb,Zhang:2019glo,Jusufi:2019ltj,Cuadros-Melgar:2020kqn}
\begin{equation}\label{OmegaR}
	\mathcal{R}_{\text {sh }}=\frac{1}{\Omega_c}=\frac{r_{\mathrm{c}}}{\sqrt{f\left(r_{\mathrm{c}}\right)}}.
\end{equation}
Using Eq.~(\ref{OmegaR}), we plot the shadow radius $\mathcal{R}_{\text{sh}}$ for different values of the parameter $\gamma$ in Fig.~\ref{omegac1}. As the parameter $\gamma$ increases, the shadow radius $\mathcal{R}_{\text{sh}}$ exhibits a monotonic increase, which is consistent with Fig.~\ref{constrain}.

\begin{figure}[!ht]
	\centering
	\subfigure[The real parts of the QNM frequencies \label{qnm2}]{
		\begin{minipage}[t]{0.5\linewidth}
			\centering
			\includegraphics[width=6cm,height=6cm]{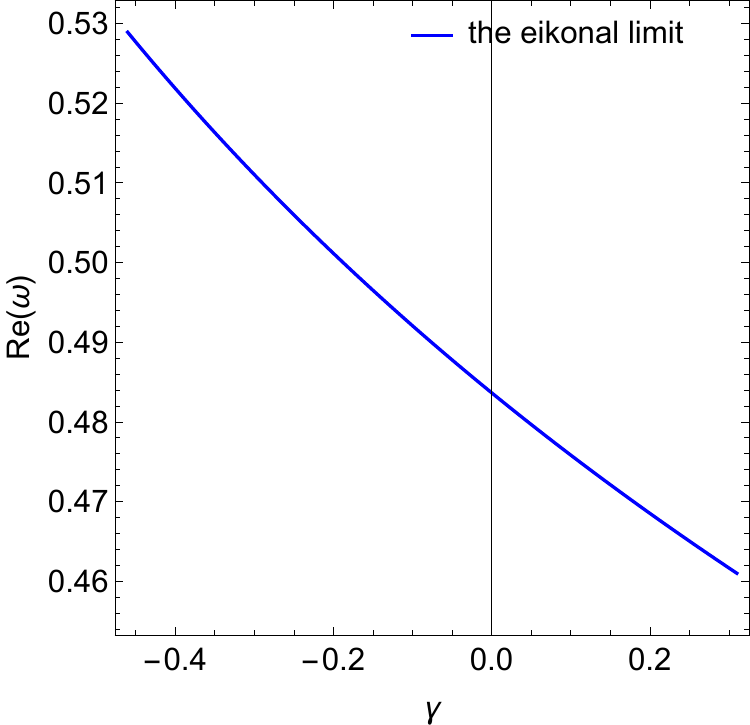}
		\end{minipage}%
	}%
	\subfigure[The negative imaginary parts of the QNM frequencies\label{qnm4}]{
		\begin{minipage}[t]{0.5\linewidth}
			\centering
			\includegraphics[width=6cm,height=6cm]{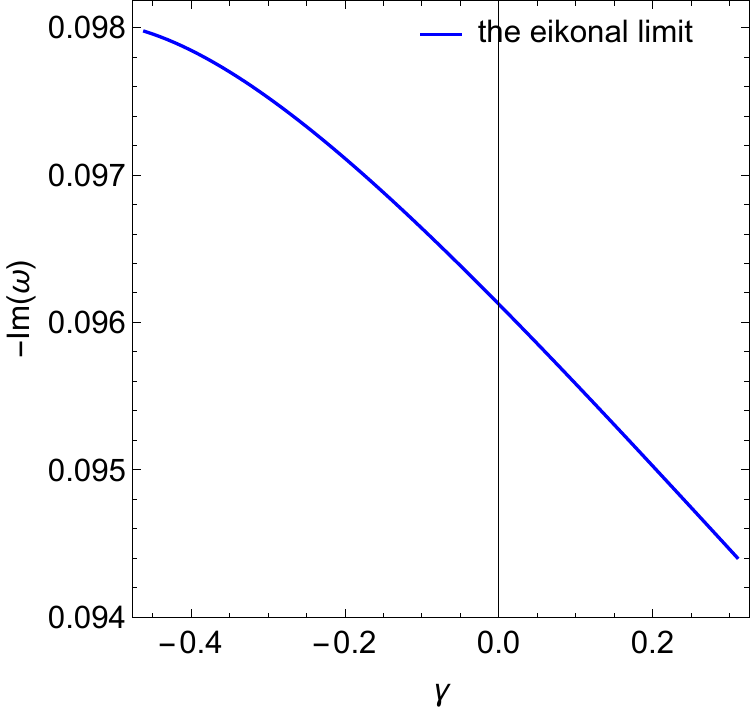}
		\end{minipage}%
	}%
	\caption{The real and negative imaginary parts of the QNM frequencies calculated via the eikonal limit. We set the overtone number $\mathfrak{n}=0$ and $M = 1$.} 
	\label{qnms2}
\end{figure}\textbf{}
\begin{figure}[!ht]
	\centering
	{
		\begin{minipage}[t]{0.5\linewidth}
			\centering
			\includegraphics[width=6cm,height=6cm]{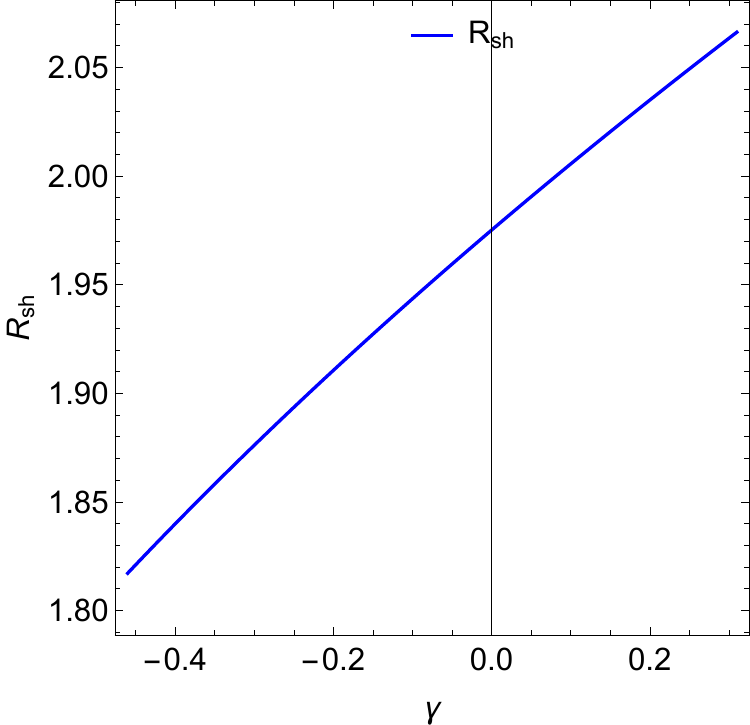}
		\end{minipage}%
	}%
	\caption{The radius $\mathcal{R}_{\text {sh }}$ of the BH shadow varies with the parameter $\gamma$. We set $M = 1$.} 
	\label{omegac1}
\end{figure}\textbf{}

\subsection{Oscillation frequencies for ${\text {M87*}}$ and  ${\text {SgrA*}}$ in Horndeski theory}\label{sec:QNMs3}

At last, we discuss the oscillation frequencies of a massless scalar perturbation field in the geometry of the supermassive BHs ${\text {M87*}}$ and ${\text {SgrA*}}$ in Horndeski theory, which, in the unit of Hertz, can be expressed as
\begin{equation}\label{31y}
	f=\frac{\operatorname{Re}(\omega)}{2 \pi M} \times \frac{c^3}{G}.
\end{equation} 
Here, $\operatorname{Re}(\omega)$ is the real part of the QNM frequencies. Taking $M_{\mathrm{M} 87^*}=6.5  \times 10^{9}M_{\odot}$, $M_{\text {SgrA*}}=4.3  \times 10^6 M_{\odot}$, and the corresponding $\operatorname{Re}(\omega)$ into Eq.~(\ref{31y}), we plot the oscillation frequencies (the fundamental mode) varying with the parameter $\gamma$ for $\mathrm{M} 87^*$ and $\mathrm{Sgr} \mathrm{A}^*$ in Fig.~\ref{fre}. It is found that the oscillation frequencies for both $\mathrm{M} 87^*$ and $\mathrm{Sgr} \mathrm{A}^*$ monotonically decrease with the parameter $\gamma$. For $\mathrm{M} 87^*$, the frequency range  is limited to $2.4\times 10^{-6}$ Hz $\lesssim f \lesssim 2.7\times 10^{-6}$ Hz due to $-0.46 \lesssim \gamma \lesssim 0.31$. And for $\mathrm{Sgr} \mathrm{A}^*$, the frequency range  is limited to $3.6\times 10^{-3}$~Hz $\lesssim f \lesssim 4.1\times 10^{-3}$~Hz due to $-0.46 \lesssim \gamma \lesssim 0.31$.

\begin{figure}[!ht]
	\centering
	\subfigure[The oscillation frequencies for $\mathrm{M} 87^*$ \label{fHz1}]{
		\begin{minipage}[t]{0.5\linewidth}
			\centering
			\includegraphics[width=6cm,height=6cm]{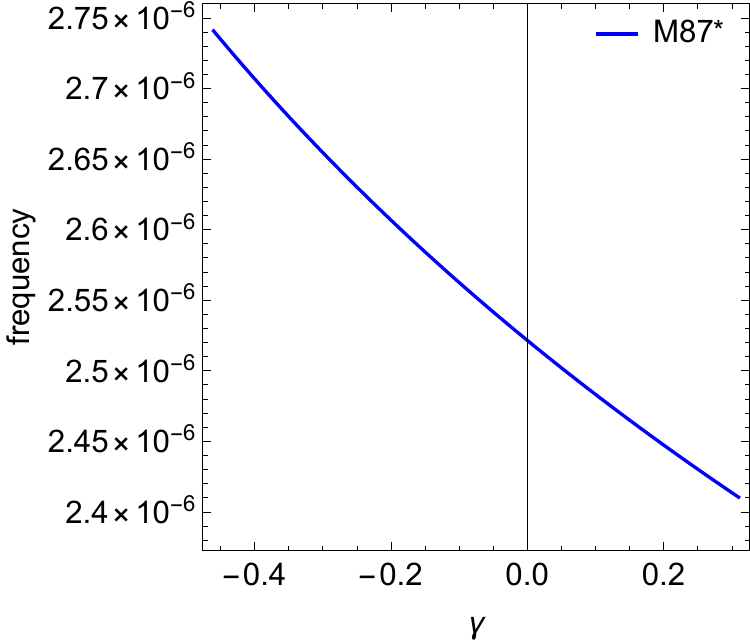}
		\end{minipage}%
	}%
	\subfigure[The oscillation frequencies for $\mathrm{Sgr} \mathrm{A}^*$ \label{fHz2}]{
		\begin{minipage}[t]{0.5\linewidth}
			\centering
			\includegraphics[width=6cm,height=6cm]{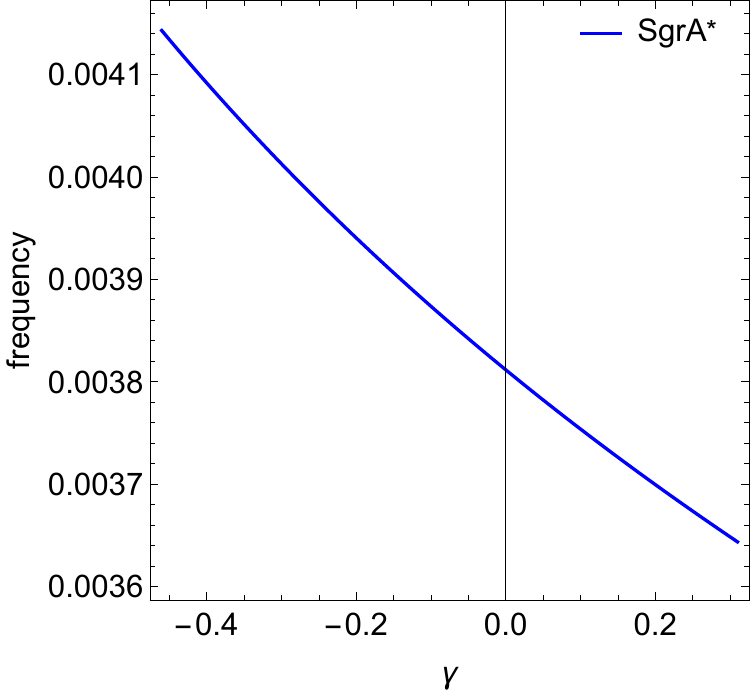}
		\end{minipage}%
	}%
	\caption{The oscillation frequencies vary with the parameter $\gamma$ for the supermassive BHs $\mathrm{M} 87^*$ and $\mathrm{Sgr} \mathrm{A}^*$. We set the overtone number $\mathfrak{n}=0$, the multipole number $\ell=2$, and $M = 1$.} 
	\label{fre}
\end{figure}\textbf{}

\section{Optical appearances of the BH in Horndeski theory}
\label{sec:4}
{In this section, we investigate the optical appearances of the BH encircled by a geometrically thin accretion disk in Horndeski theory.} We analyze the impact of the parameter $\gamma$ on these optical appearances.

\subsection{\textbf{Direct emission, photon ring, and lensing ring}}
\label{sec:4-1}
To investigate the optical appearances of the BH in Horndeski theory, we suppose that the equatorial plane of the BH is fixed. The static observer and the light source are positioned at infinity in the north pole and south pole directions, respectively. We define the total number of photon orbits around the BH as $\boldsymbol{N} \equiv \phi / (2 \pi)$, where $\phi$ is the total deflection angle of the photon given by Eq.~(\ref{eq19y}). Based on the value of $\boldsymbol{N}$, one can classify the light rays into the following three categories~\cite{Gralla:2019xty}:
\begin{itemize}
	\item \emph{Direct Emission:}~~ $\boldsymbol{N} < 3/4$. Light rays and the equatorial plane intersect at most once;
	\item \emph{Lensing Ring:}~~ $3/4 < \boldsymbol{N} < 5/4$. Light rays and the equatorial plane intersect twice;
	\item \emph{Photon Ring:}~~ $\boldsymbol{N} > 5/4$. Light rays and the equatorial plane intersect at least three times.
\end{itemize}

For the BH in Horndeski theory, the total number $\boldsymbol{N}$  of photon orbits is only dependent on the impact parameter $b/M$ and the parameter $\gamma$. We plot the relationship between the total number $\boldsymbol{N}$ of photon orbits and the impact parameter $b/M$ for different values of the parameter $\gamma$ (= -0.3, 0, 0.3), as shown in Fig.~\ref{transfer1}. The graph is divided into three regions: ``Direct Emission", ``Lensing Ring", and ``Photon Ring" corresponding to the three categories of light rays mentioned above. These regions are separated by two horizontal dashed lines at $\boldsymbol{N} = 3 / 4$ and $\boldsymbol{N} = 5 / 4$. In Tab~\ref{tab:my_table}, we present the ranges of the impact parameter corresponding to the direct emission, lensing ring, and photon ring, respectively. For the three cases of $\gamma$, the evolution of $\boldsymbol{N}$ with respect to the impact parameter $b/M$ is consistent, i.e., $\boldsymbol{N}$ increases with $b/M$, reaches a peak, and then decreases. Comparing the peaks corresponding to different values of the parameter $\gamma$, we find that the peak point shifts to the right along the horizontal axis as the parameter $\gamma$ increases. This indicates that the size of the photon ring increases with the parameter $\gamma$.

\begin{table}[!ht]
	\centering
	\begin{tabular}{cccc}
		\hline
		Parameter $\gamma$ & Direct Emission & Lensing Ring & Photon Ring \\
		\hline
		-0.3 & $b / M \notin(4.720,5.948)$ & $b / M \in(4.720,4.908)$ or $b / M \in(4.950,5.948)$ & $b / M \in(4.908,4.950)$ \\
		0 & $b / M \notin(5.014,6.181)$ & $b / M \in(5.014,5.187)$ or $b / M \in(5.231,6.181)$ & $b / M \in(5.187,5.231)$ \\
		0.3 & $b / M \notin(5.272,6.422)$ & $b / M \in(5.272,5.435)$ or $b / M \in(5.461,6.422)$ & $b / M \in(5.435,5.461)$ \\
		\hline
	\end{tabular}
	\caption{The ranges of the impact parameter $b/M$ corresponding to the direct emission, lensing ring, and photon ring, respectively. We set $\gamma$= -0.3, 0, 0.3 and $M=1$.}
	\label{tab:my_table}
\end{table}
\begin{figure}[!ht]
	\centering
	{
		\begin{minipage}[t]{0.5\linewidth}
			\centering
			\includegraphics[width=6cm,height=6cm]{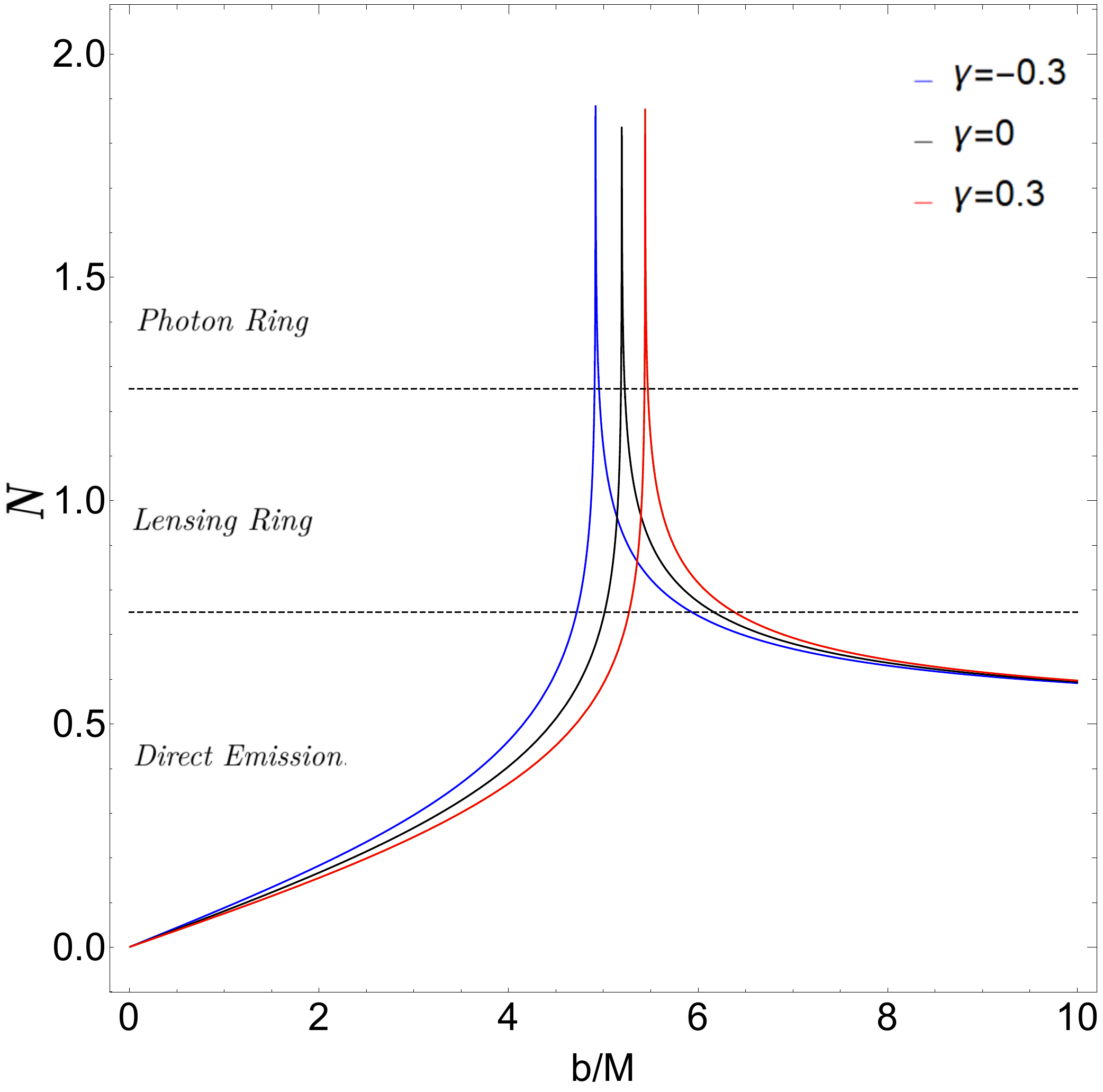}
		\end{minipage}%
	}%
	\caption{The relationship between the total number $\boldsymbol{N}$ of photon orbits and the impact parameter $b / M$ for the BH in Horndeski theory. The blue, black, and red lines correspond to $\gamma = -0.3$, $\gamma = 0$, and $\gamma = 0.3$, respectively. We set $M=1$.}
	\label{transfer1}
\end{figure}

Based on the above classification of light rays, we plot the trajectories of photons for different values of the parameter $\gamma$ in polar coordinates $(r, \phi)$ (see Fig.~\ref{tra}). In the figure, the black disk represents the BH in Horndeski theory, the red ring indicates the photon sphere, and the vertical black line passing through the origin denotes the equatorial plane. The observer is positioned at infinity in the north polar direction, corresponding to the right side of the panel. Taking $\gamma = -0.3$ as an example (see Fig.~\ref{trajI1}), one can find that for the observer, the direct emission occupies a significant portion of the observed appearances of the BH (see the red lines). In the figure, the direct emission can be divided into upper and lower sections based on the value of the impact parameter $b / M$. In the upper section ($1 / 4 < \boldsymbol{N} < 3 / 4$), photons with larger impact parameters intersect the equatorial plane only once and then escape the BH. In the lower section ($\boldsymbol{N} < 1 / 4$), photons with smaller impact parameters fall into the BH. The lensing ring occupies a smaller central region of the observed appearances of the BH (see the blue lines). For the lensing ring, all photons intersect the equatorial plane twice, and the distance between the two intersections increases with the impact parameter. Similarly, the lensing ring could also be divided into upper and lower sections based on the value of the impact parameter $b / M$. The upper section (photons with larger impact parameters) could escape the BH, while the lower section (photons with smaller impact parameters) will fall into the BH. As for the photon ring, it occupies only a very narrow region of the observed appearance of the BH (see the green lines). The corresponding photons intersect the equatorial plane more than three times. Comparing the three cases corresponding to different $\gamma$'s, it is observed that the trajectories of photons in the three pictures exhibit similarity. The influence of $\gamma$ on the trajectories of photons is visually evident in the radii of the black disk and the photon sphere. Both of them expand as $\gamma$ increases (see Figs.~\ref{trajI1}, ~\ref{trajII1}, and ~\ref{trajIII1}, click to zoom in).

 \begin{figure}[!ht]
	\centering
	\subfigure[$\gamma=-0.3$\label{trajI1}]{
		\begin{minipage}[t]{0.33\linewidth}
			\centering
			\includegraphics[width=4cm,height=4cm]{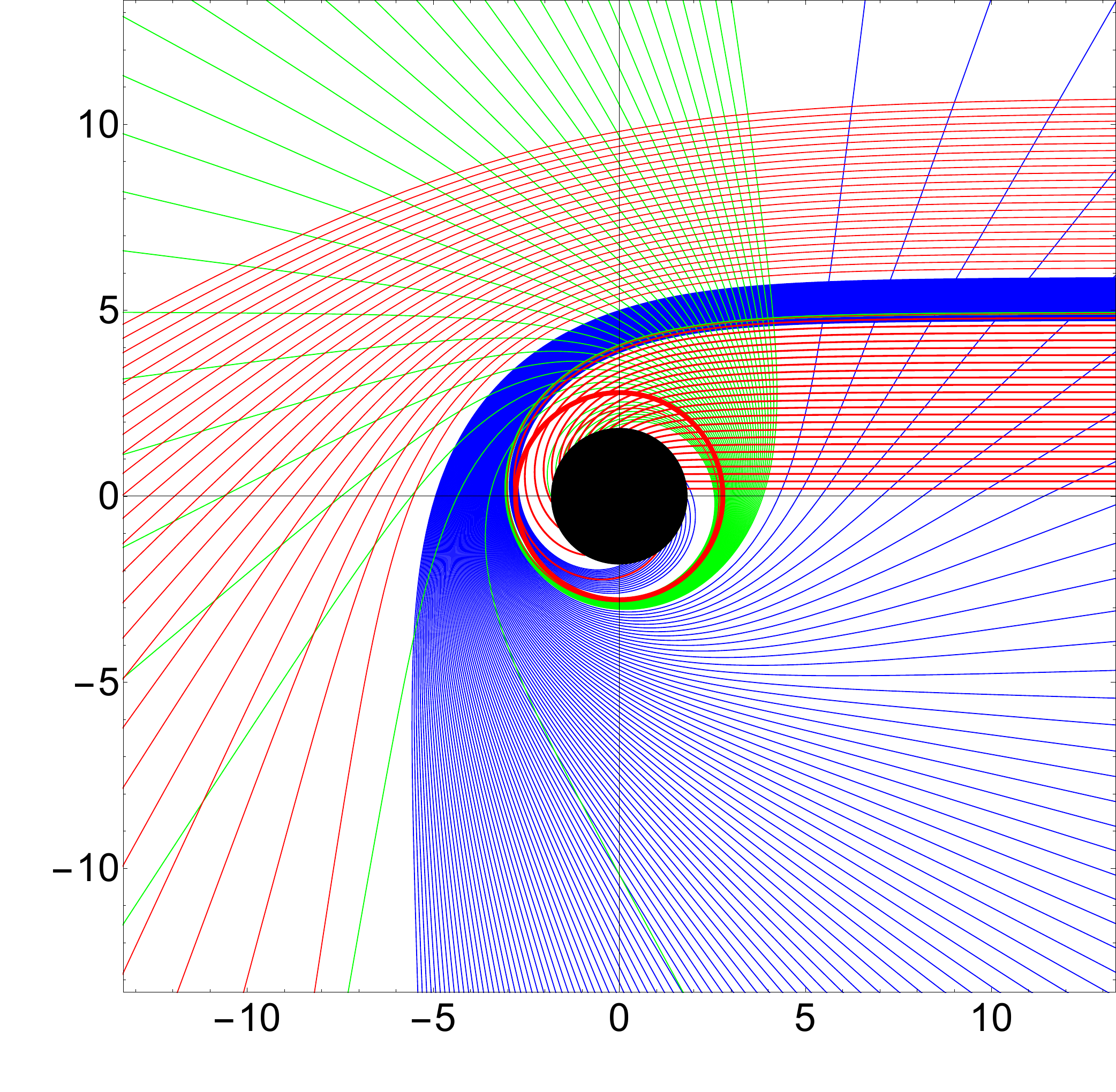}
		\end{minipage}%
	}%
	\subfigure[$\gamma=0$\label{trajII1}]{
		\begin{minipage}[t]{0.33\linewidth}
			\centering
			\includegraphics[width=4cm,height=4cm]{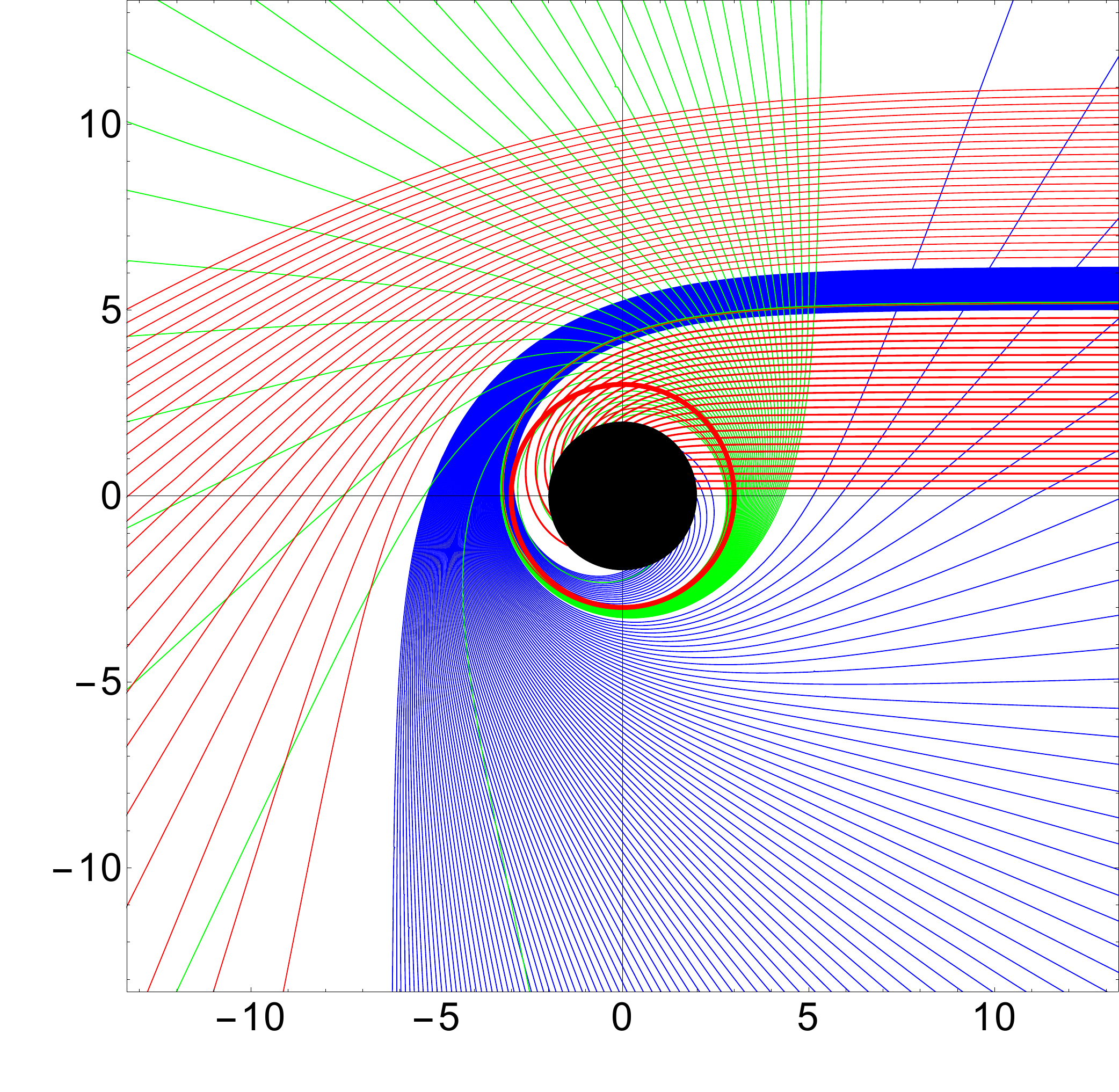}
		\end{minipage}%
	}%
	\subfigure[$\gamma=0.3$\label{trajIII1}]{
		\begin{minipage}[t]{0.33\linewidth}
			\centering
			\includegraphics[width=4cm,height=4cm]{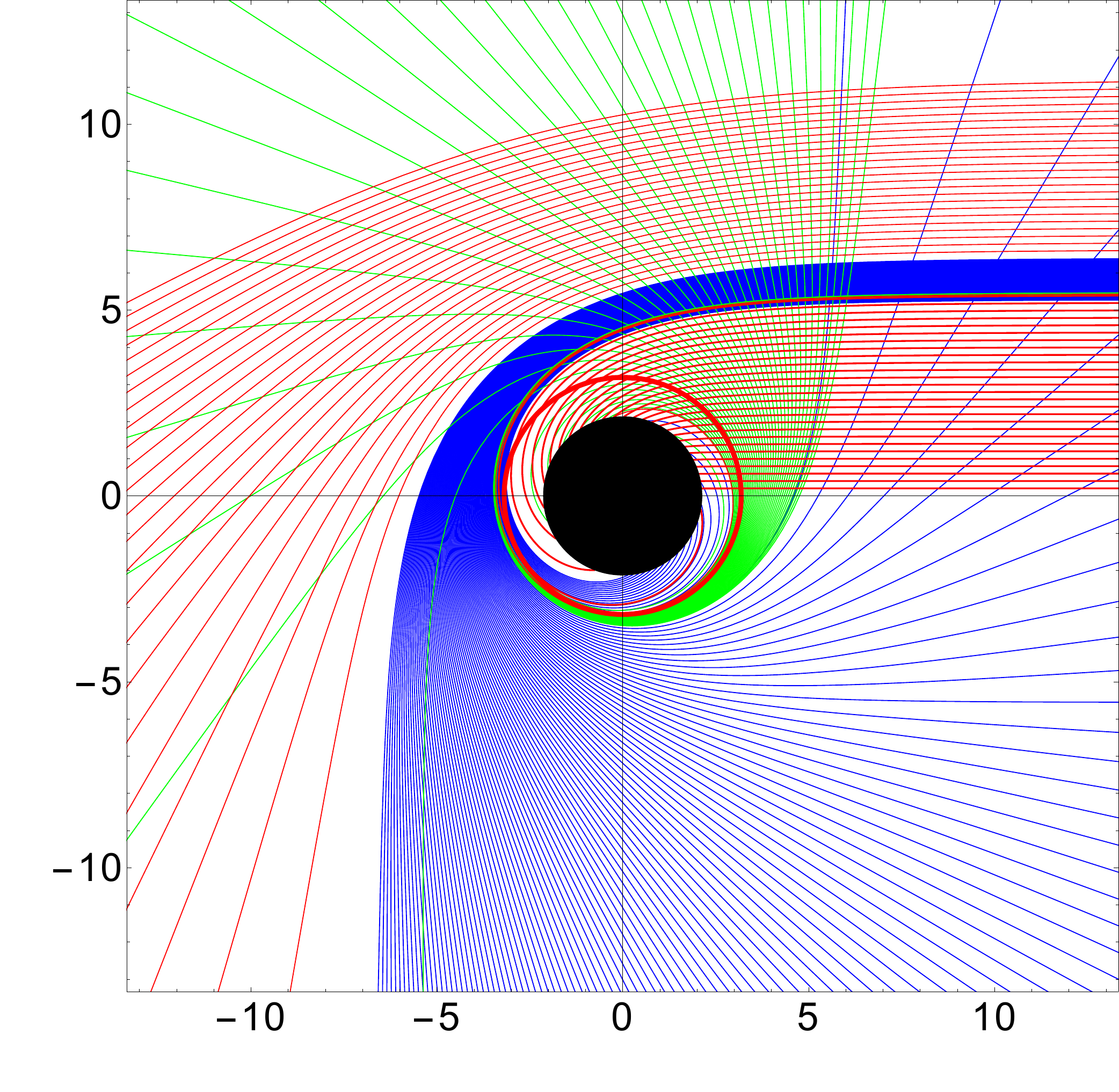}
		\end{minipage}
	}%
	\caption{The trajectories of photons for different values of the parameter $\gamma$ in polar coordinates $(r, \phi)$. The red lines, blue lines and green lines represent the direct emission, lensing ring, and photon ring, respectively.  We set $\gamma$ = -0.3, 0, 0.3 and $M = 1$. The spacing of the impact parameter for the direct emission, lensing ring, and photon ring is set to $1 / 5$, $1 / 100,$ and $1 / 1000$, respectively.}
	\label{tra}
\end{figure}

\subsection{\textbf{Transfer functions}}
\label{sec:transfer}

Now, we study the transfer functions of the BH surrounded by a geometrically thin accretion disk in Horndeski theory. The disk is still located on the equatorial plane and emits radiation isotropically for a static observer at infinity in the north pole. The emitted specific intensity of the disk can be expressed as
\begin{equation}
	I_\nu^{\mathrm{em}}=I(r),
\end{equation}
where $\nu$ represents the emission frequency in the static farme. Considering that $I_\nu / \nu^3$ remains conserved along a ray, the specific intensity of the radiation emitted from a radius $r$ and received at any frequency $\nu'$ can be given by~\cite{Gralla:2019xty}
\begin{equation}
	I_{\nu^{\prime}}^{\text {obs }}=f(r)^{3/2} I(r),
\end{equation}
where $f(r)=1-\frac{2M}{r}-\frac{\gamma}{r^2}$. Then, we can obtain the total observed specific intensity as follows~\cite{Gralla:2019xty}:
\begin{equation}
	I^{\mathrm{obs}}=\int I_{\nu^{\prime}}^{\text {obs }} \mathrm{d} \nu^{\prime}=\int f(r)^2 I_\nu^{\mathrm{em}} \mathrm{~d} \nu=f(r)^2 I(r).
\end{equation}
Since we are considering a geometrically thin accretion disk located on the equatorial plane, each time a photon intersects this plane, it acquires brightness from the disk's emission, as shown in Fig.~\ref{tra}. Consequently, the observed intensity is given by~\cite{Gralla:2019xty}
\begin{equation}\label{observedintensity}
	I^{\text {obs }}(b)=\left.\sum_n f(r)^2 I\right|_{r=r_n(b)},
\end{equation}
which a sum of the intensities from each intersection. Here, $r_n(b) \ (n = 1, 2, 3, \ldots)$ is the transfer function, which represents the radial coordinate of the $n$-th intersection point between the photon and the geometrically thin accretion disk. The slope of the function $r_n(b)$ at each $b$ yields the demagnification factor at the point, denoted as $dr/db$. In Fig.~\ref{transfer2}, we plot the first three transfer functions for different values of the parameter $\gamma$. One can find that for a given $\gamma$, there exists a critical impact parameter. When an impact parameter is less than the critical impact parameter, the transfer functions do not support it. Taking $\gamma = -0.3$ as an example (see Fig.~\ref{transferI}), one can observe that the transfer functions do not support $b \lesssim 2.63 M$. This critical value is greater than the event horizon radius ($r_{\text{h}} \approx 1.84 M$), but smaller than the photon sphere radius ($r_{\text{ph}} \approx 2.78 M$). The first transfer function with a slope close to 1 $(n=1)$ corresponds to the direct emission  (see the red line in Fig.~\ref{transferI}), which represents the redshifted source profile of the geometrically thin accretion disk. The second transfer function with a steeper slope $(n=2)$ pertains to the lensed ring (see the blue line in Fig.~\ref{transferI}), which, for the observer, manifests a highly demagnified image of the back side of the disk. The third transfer function with the steepest slope $(n=3)$ is associated with the photon ring (see the green line in Fig.~\ref{transferI}), which coresponds to an extremely demagnified image of the front side of the disk. This implies that the first transfer function significantly contributes to the total flux, while the second transfer function contributes to a lesser extent, and the third transfer function makes very little contribution to the total flux. For $\gamma = 0$ (see Fig.~\ref{transferII}) and $\gamma = 0.3$ (see Fig.~\ref{transferIII}), the transfer functions exhibit similarities to the case when $\gamma = -0.3$. However, a larger value of the parameter $\gamma$ results in an increased initial $b/M$ value for the transfer functions. But, when the parameter $\gamma$ is larger, the transfer function has an increased critical impact parameter.

\begin{figure}[!ht]
	\centering
	\subfigure[$\gamma=-0.3$\label{transferI}]{
		\begin{minipage}[t]{0.33\linewidth}
			\centering
			\includegraphics[width=4cm,height=4cm]{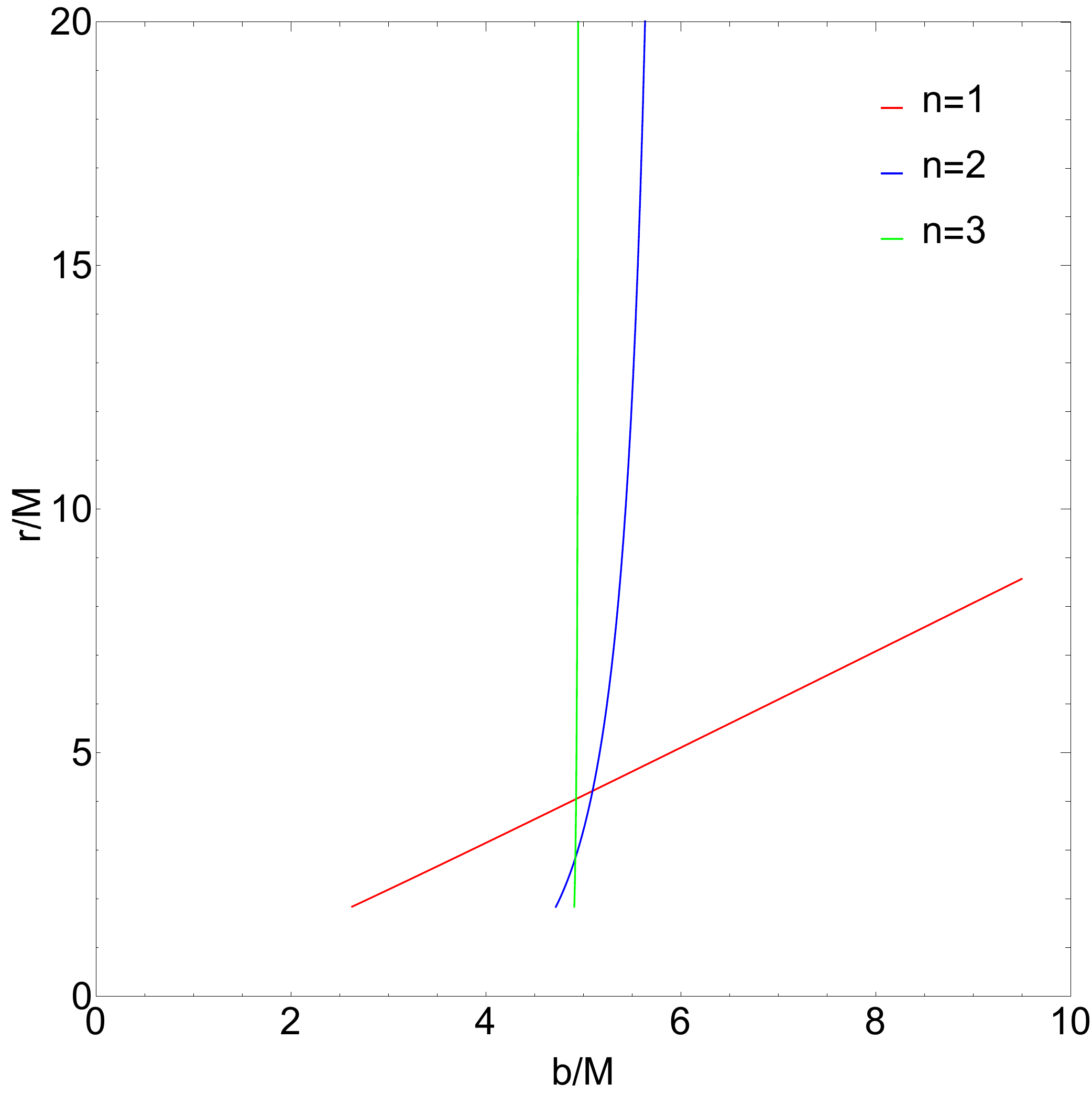}
		\end{minipage}%
	}%
	\subfigure[$\gamma=0$\label{transferII}]{
		\begin{minipage}[t]{0.33\linewidth}
			\centering
			\includegraphics[width=4cm,height=4cm]{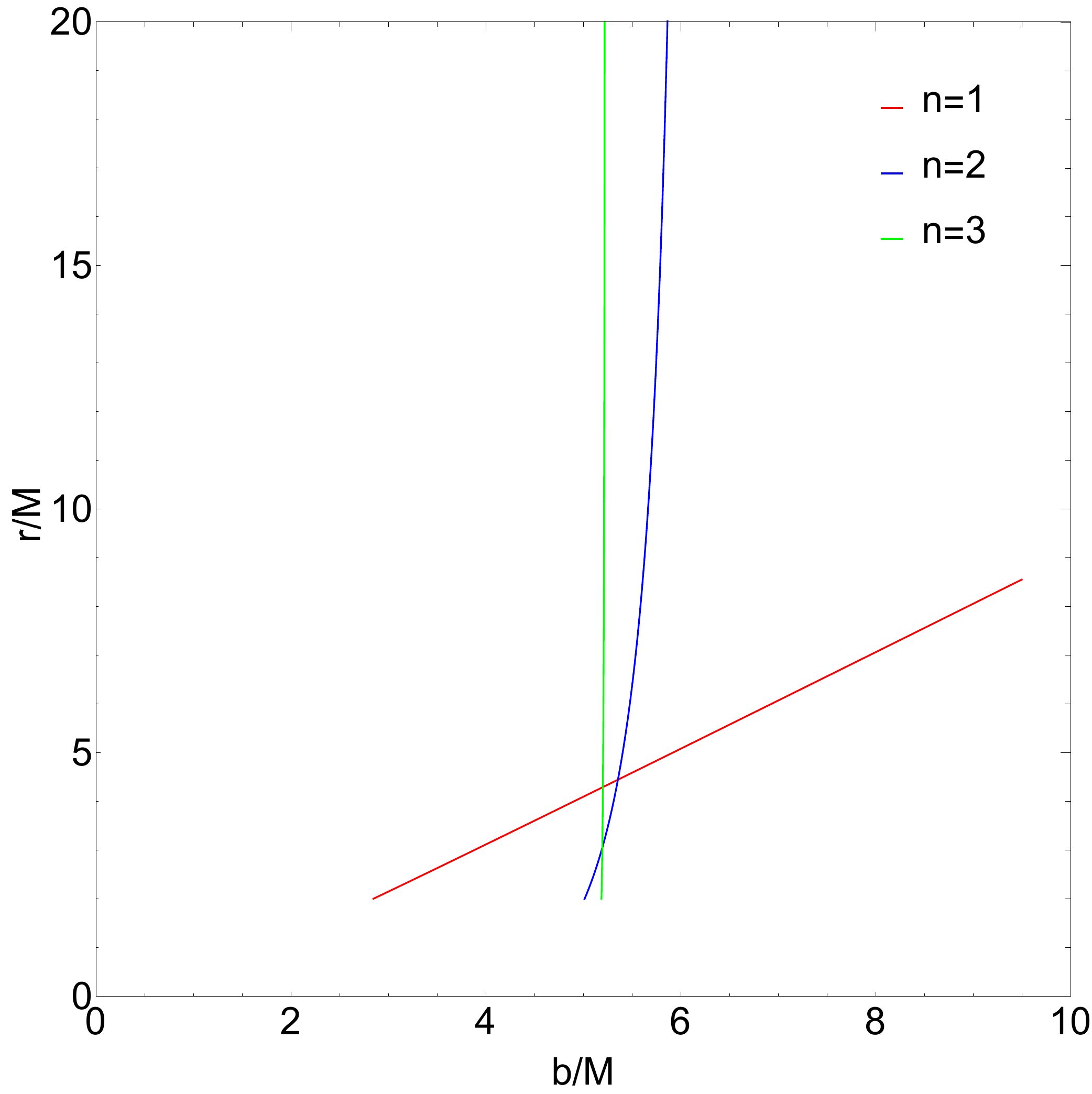}
		\end{minipage}%
	}%
	\subfigure[$\gamma=0.3$\label{transferIII}]{
		\begin{minipage}[t]{0.33\linewidth}
			\centering
			\includegraphics[width=4cm,height=4cm]{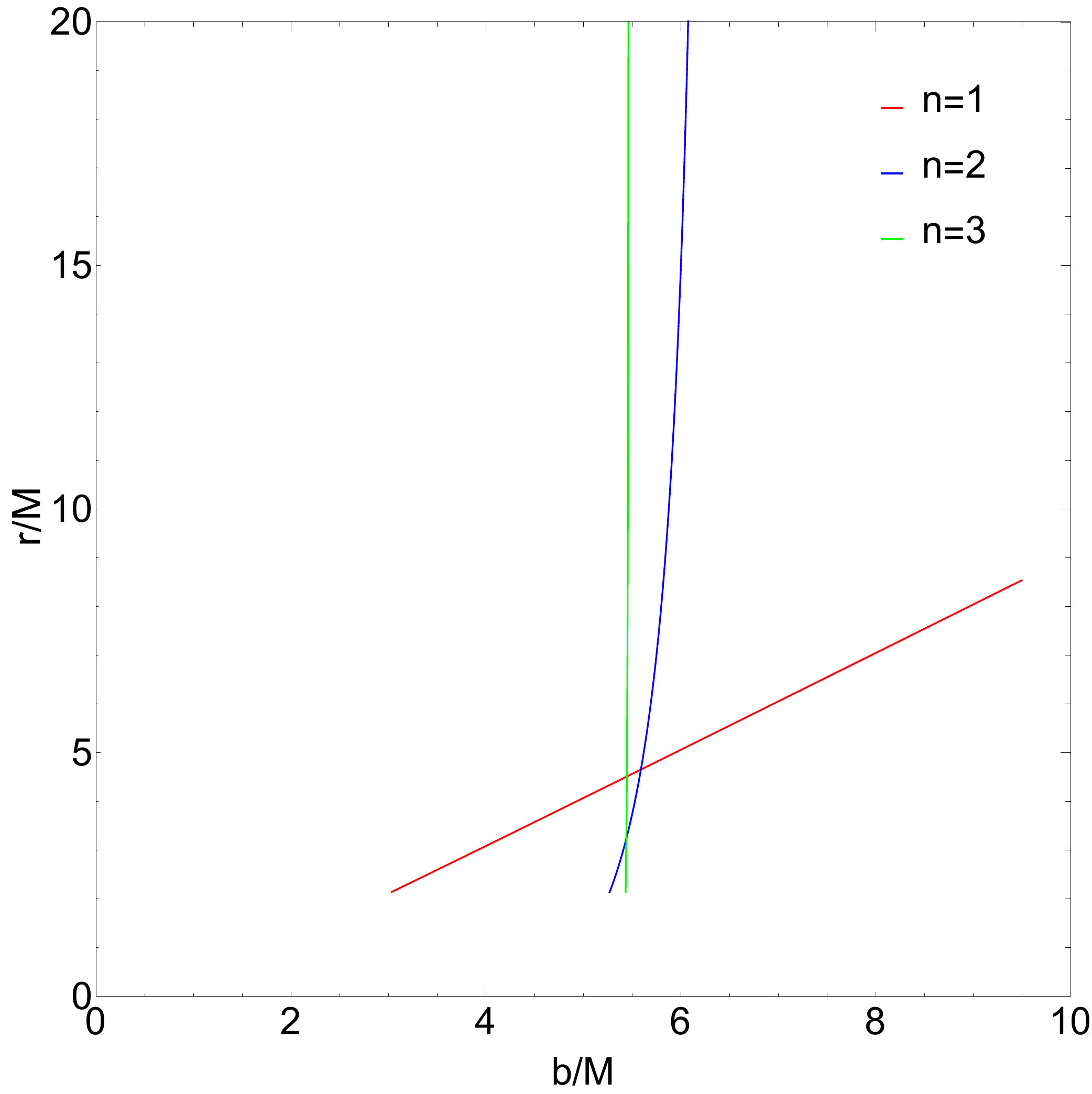}
		\end{minipage}
	}%
	\caption{The first three transfer functions of the BH in Horndeski theory for different values of the parameter $\gamma$. The red, blue, and green lines represent the radial coordinates of photons at their first, second, and third intersections with the geometrically thin accretion disk, respectively. We set $\gamma=-0.3,0,0.3$ and $M=1$.}
	\label{transfer2}
\end{figure}

\subsection{\textbf{Optical appearances for three emission models}}
\label{sec:4-3}
In this section, we investigate the optical appearances of the BH in Horndeski theory using three specific emission models of  geometrically thin accretion disks, as illustrated in Fig.~\ref{modela}. 

The first emission model can be expressed as
\begin{equation}\label{emissionmodel1}
	I^{\mathrm{em}}_1(r)= \begin{cases}I_0\left[\frac{1}{r-\left(r_{\mathrm{isco}}-1\right)}\right]^2, & r>r_{\mathrm{isco}}, \\ 0, & r \leq r_{\mathrm{isco}}.\end{cases}
\end{equation}
Here, we assume that the emission starts from the innermost stable circular orbit $r_{\mathrm{isco}}$. Then, it suddenly increases, reaches a peak, and rapidly declines, as shown in Fig.~\ref{model1}. 

The second emission model assumes that the emission starts from the event horizon and decays towards the innermost stable circular orbit at a slower rate compared to the first emission model. As shown in Fig.~\ref{model2}, the emitted specific intensity of the model is given by
\begin{equation}\label{emissionmodel2}
	I^{\mathrm{em}}_2(r)= \begin{cases}I_0 \frac{\frac{\pi}{2}-\arctan \left[r-\left(r_{\mathrm{isco}}-1\right)\right]}{\frac{\pi}{2}-\arctan \left[r_h-\left(r_{\mathrm{isco}}-1\right)\right]}, & r>r_h, \\ 0, & r \leq r_h .\end{cases}
\end{equation}

In the third emission model, the emission is assumed to occur at the photon sphere radius $r_{\mathrm{ph}}$ and then rapidly decay at a cubic rate, as shown in Fig.~\ref{model3}. The emitted specific intensity is given by
\begin{equation}\label{emissionmodel3}
	I^{\mathrm{em}}_3(r)= \begin{cases}I_0\left[\frac{1}{r-\left(r_{\mathrm{ph}}-1\right)}\right]^3, & r>r_{\mathrm{ph}}, \\ 0, & r \leq r_{\mathrm{ph}}.\end{cases}
\end{equation}
\begin{figure}[!ht]
	\centering
	\subfigure[First emission model \label{model1}]{
		\begin{minipage}[t]{0.33\linewidth}
			\centering
			\includegraphics[width=4cm,height=4cm]{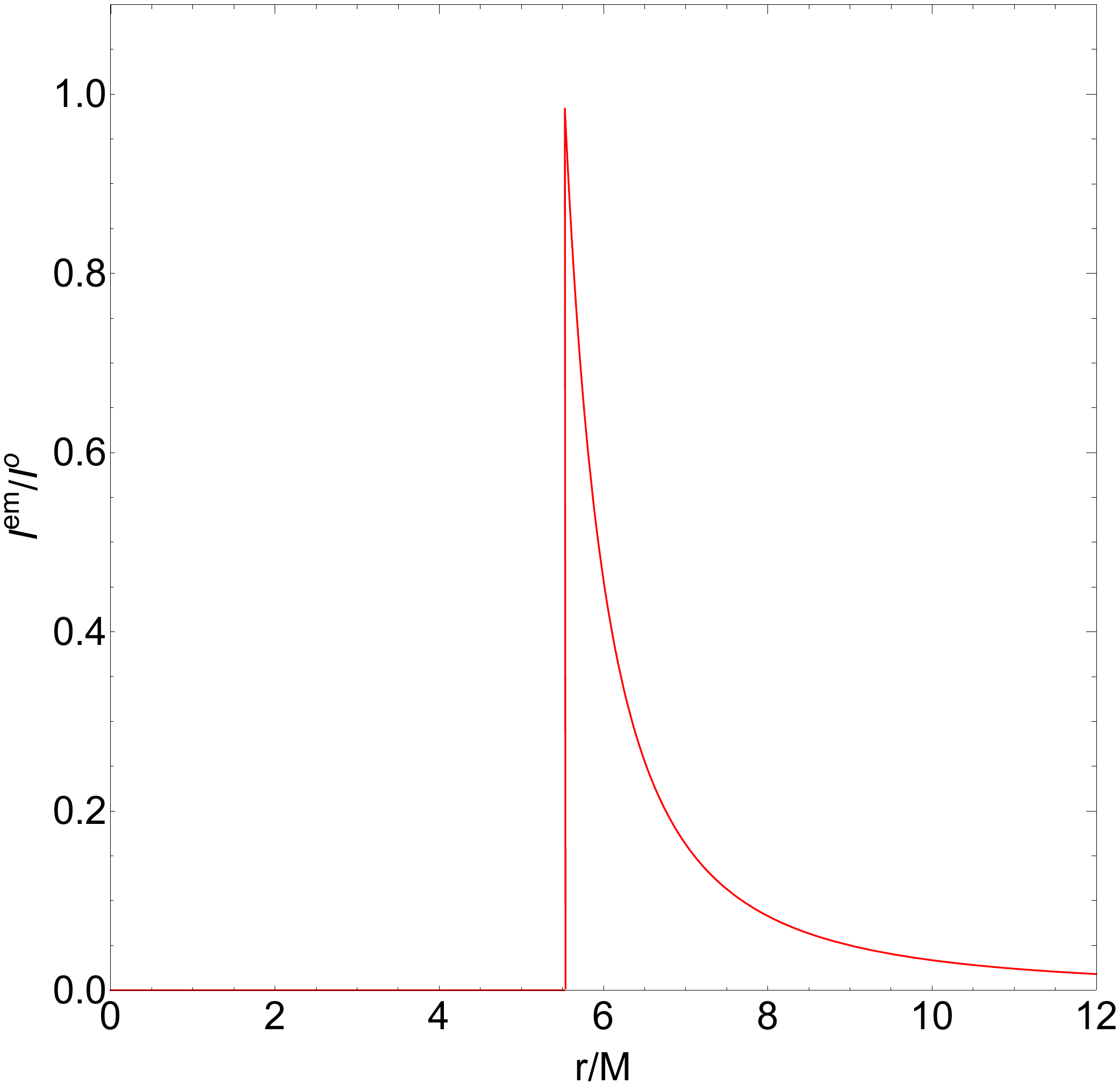}
		\end{minipage}%
	}%
	\subfigure[Second emission model \label{model2}]{
		\begin{minipage}[t]{0.33\linewidth}
			\centering
			\includegraphics[width=4cm,height=4cm]{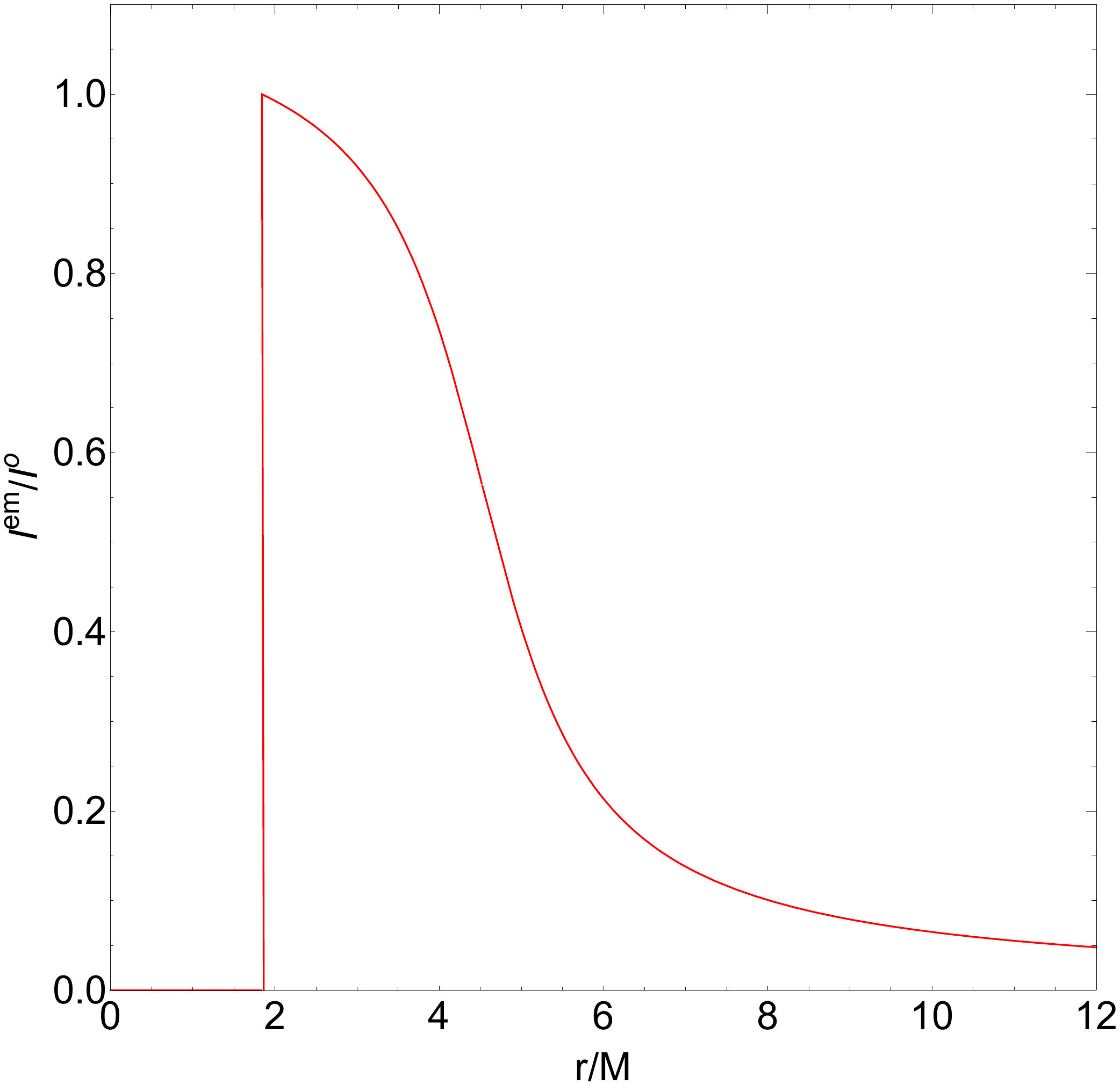}
		\end{minipage}%
	}
	\subfigure[Third emission model \label{model3}]{
		\begin{minipage}[t]{0.33\linewidth}
			\centering
			\includegraphics[width=4cm,height=4cm]{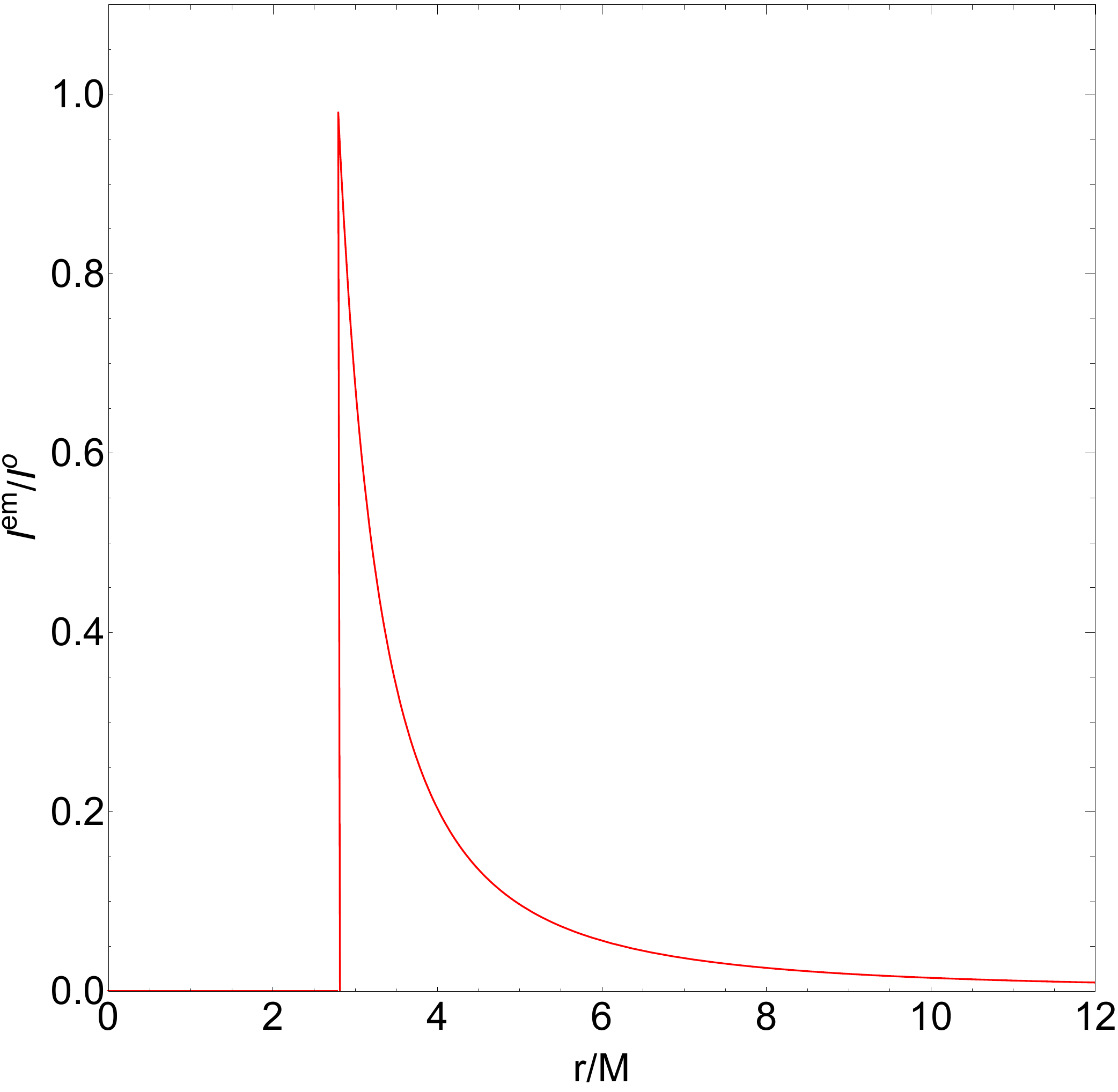}
		\end{minipage}%
	}
	\caption{The emitted specific intensites of geometrically thin accretion disks for three emission models.}
	\label{modela}
\end{figure}

Combining the observed intensity~(\ref{observedintensity}) and the three emission models~(\ref{emissionmodel1}-\ref{emissionmodel3}), we plot the optical appearances of the BH in Horndeski theory with $\gamma = -0.3$, see Fig.~\ref{densityplot1}. For other valid values of $\gamma$ (e.g., $\gamma = 0$ and $\gamma = 0.3$), it is challenging to make intuitive comparisons between their results and the results obtained for $\gamma = -0.3$. Therefore, we will not present additional results here and focus solely on $\gamma = -0.3$ as an example for further discussion. 

For the first emission model, the observed specific intensity has a peak at $b / M \approx 6.43$ (see Figs.~\ref{Iobdensity1}), corresponding to the direct emission. At $b / M \approx 5.24$, there is a narrow and slightly lower peak, which corresponds to the lensing ring. At the far left of Fig.~\ref{Iobdensity1}, there is an extremely narrow peak at $b / M \approx 4.95$, which corresponds to the photon ring. Based on the first emission model, we present the two-dimensional image of the BH as observed by the observer in Fig.~\ref{Iobdensity4}. It is found that there is a bright ring that gradually fades in intensity as it extends outward from the outer edge of the dark region, which represents the direct emission. Since the demagnification factor for the first transfer function corresponding to the direct emission is close to $1$, it contributes the majority of the total flux. In the dark region of the two-dimensional image, there is a narrow bright ring representing the lensing ring. The demagnification factor for the second transfer function corresponding to the lensing ring is relatively large, so it contributes a smaller portion of the total flux. Moreover, clicking to zoom in Fig.~\ref{Iobdensity4}, at the innermost part of the two-dimensional image, one can see that there is a very narrow and faint ring (i.e., the photon ring) near the lensing ring. The demagnification factor for the third transfer function corresponding to the photon ring is extremely large. As a result, it makes a minimal contribution to the total flux, causing the photon ring to appear less prominent in the two-dimensional image.

The observed specific intensity for the second emission model is shown in Fig.~\ref{Iobdensity2}. We find that the direct emission appears at $b / M \approx 2.66$, and the lensed ring overlaps with the direct emission at $b / M \approx 4.74$. From the two-dimensional image of the BH shown in Fig.~\ref{Iobdensity5}, we can also clearly see that the brightness of the two-dimensional image appears at $b / M \approx 2.66$ and increases with the radius. 
  
The observed specific intensity for the third emission model is shown in Fig.~\ref{Iobdensity3}. We find that the direct emission appears at $b / M \approx 3.61$, and the lensing ring appears at $b / M \approx 4.94$. The lensing ring has a higher peak intensity compared to the direct emission. As seen in Fig.~\ref{Iobdensity6}, the brightness of the two-dimensional image suddenly appears at $b / M \approx 3.61$ and decreases with the radius. Then, the bright light appears again at $b / M \approx 4.94$ and decreases with the radius again. Compared to the first and second emission models, the emitted specific intensity of the third emission model (which starts at the event horizon) results in a dimmer two-dimensional image due to the redshift effect.

\begin{figure}[!ht]
	\centering
	\subfigure[\label{Iobdensity1}]{
		\begin{minipage}[t]{0.33\linewidth}
			\centering
			\includegraphics[width=4cm,height=4cm]{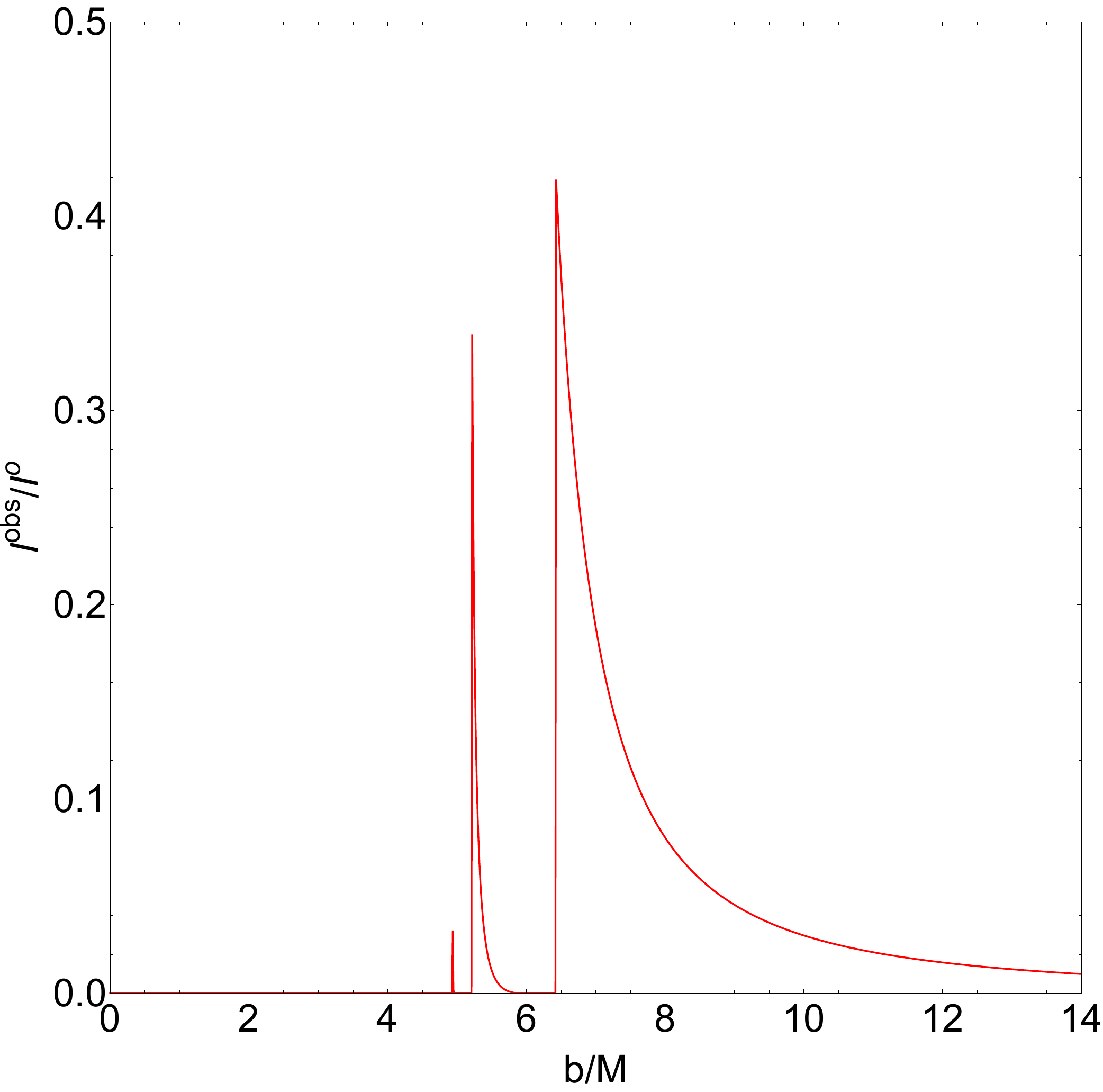}
		\end{minipage}%
	}%
	\subfigure[\label{Iobdensity2}]{
		\begin{minipage}[t]{0.33\linewidth}
			\centering
			\includegraphics[width=4cm,height=4cm]{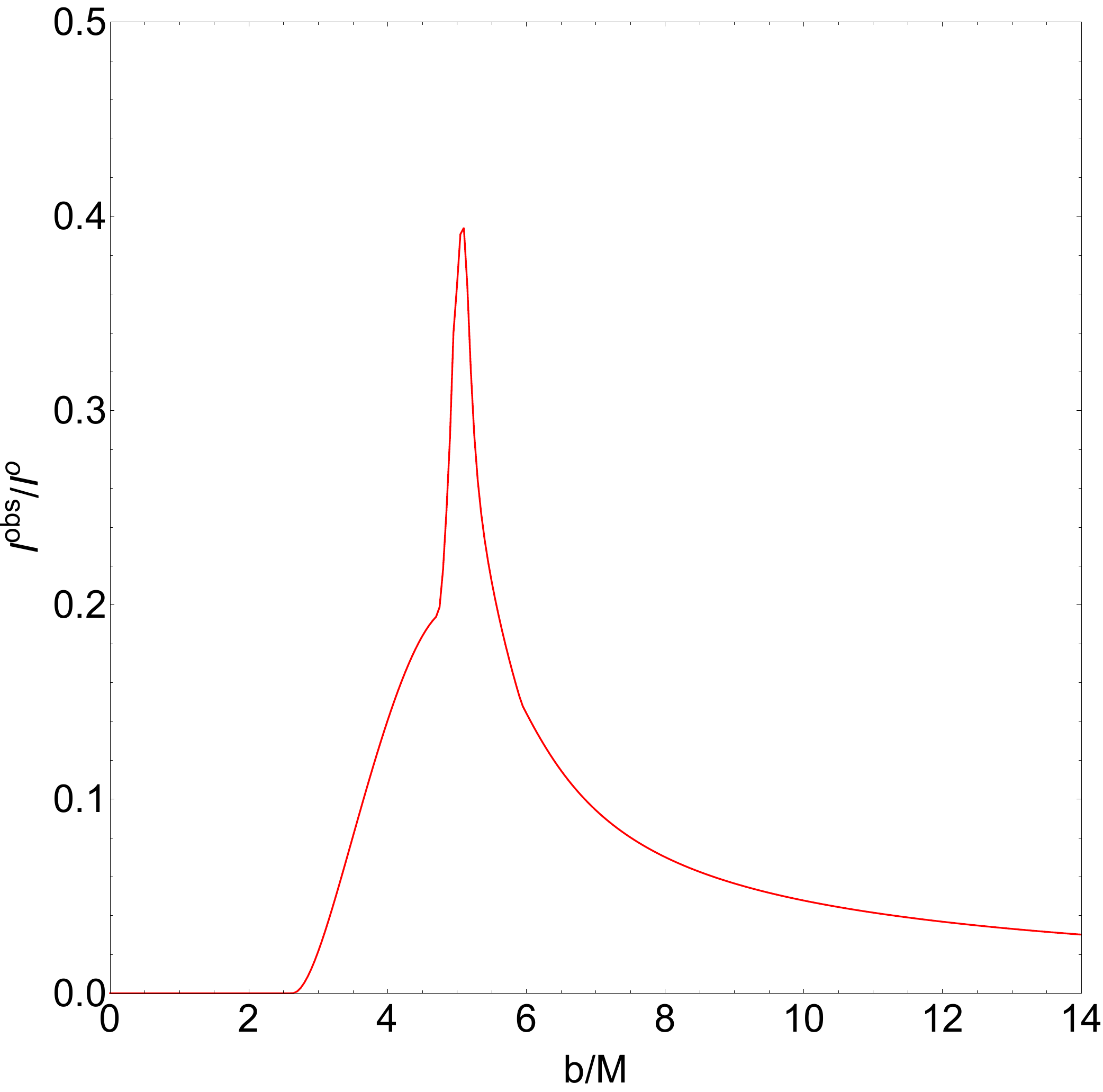}
		\end{minipage}%
	}%
	\subfigure[\label{Iobdensity3}]{
		\begin{minipage}[t]{0.33\linewidth}
			\centering
			\includegraphics[width=4cm,height=4cm]{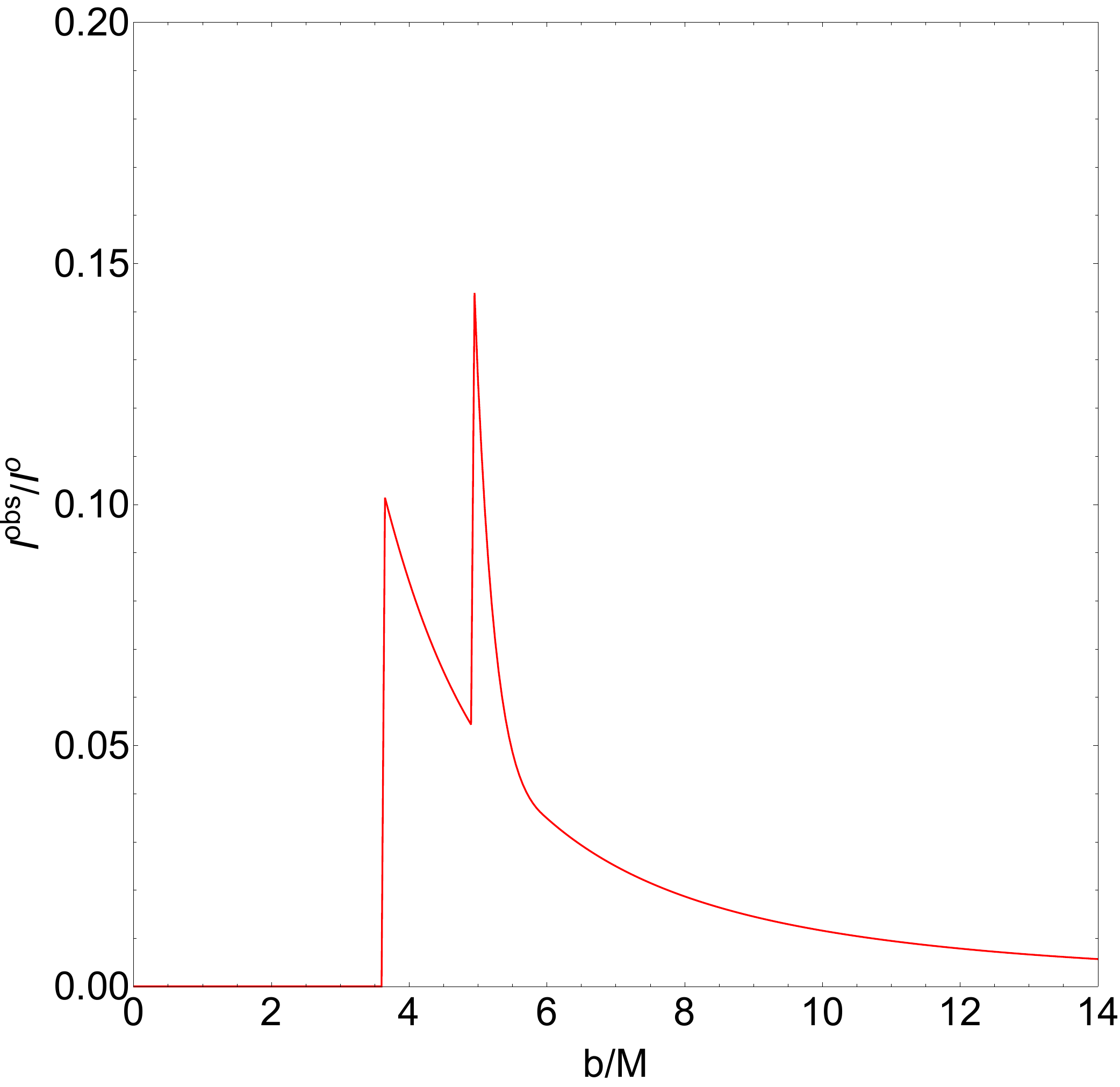}
		\end{minipage}
	}%
	\quad
	\subfigure[\label{Iobdensity4}]{
		\begin{minipage}[t]{0.32\linewidth}
			\centering
			\includegraphics[width=4cm,height=4cm]{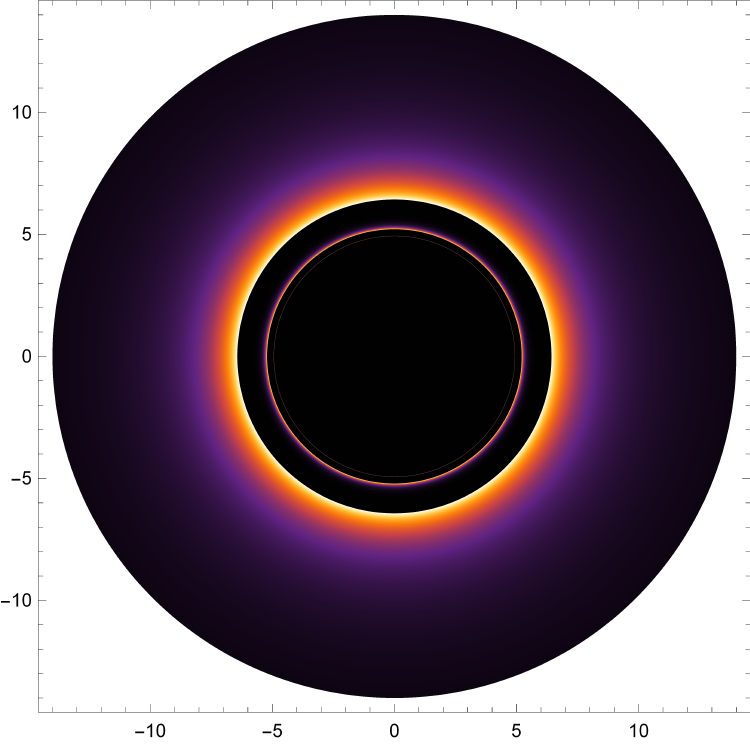}
		\end{minipage}
	}%
	\subfigure[\label{Iobdensity5}]{
		\begin{minipage}[t]{0.32\linewidth}
			\centering
			\includegraphics[width=4cm,height=4cm]{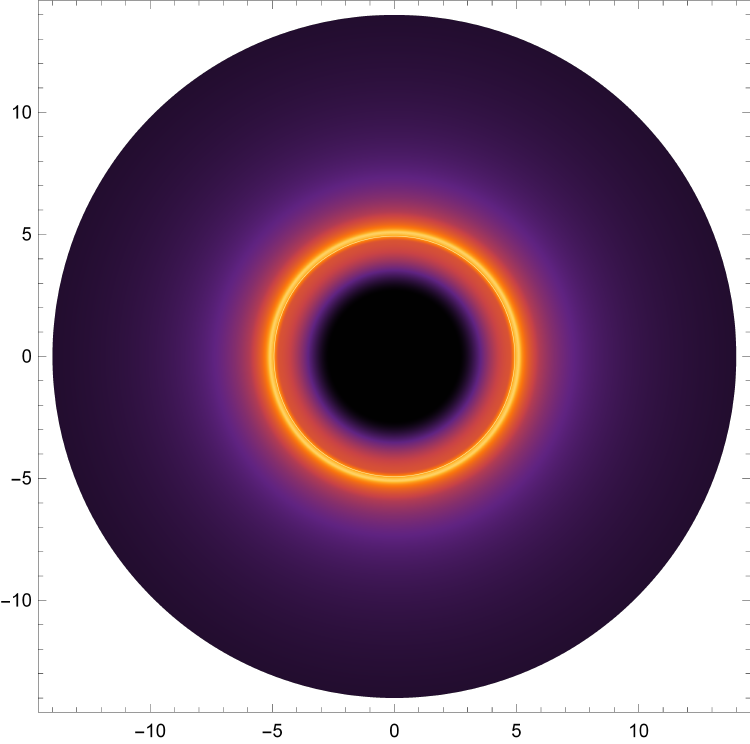}
		\end{minipage}
	}%
	\subfigure[\label{Iobdensity6}]{
		\begin{minipage}[t]{0.32\linewidth}
			\centering
			\includegraphics[width=4cm,height=4cm]{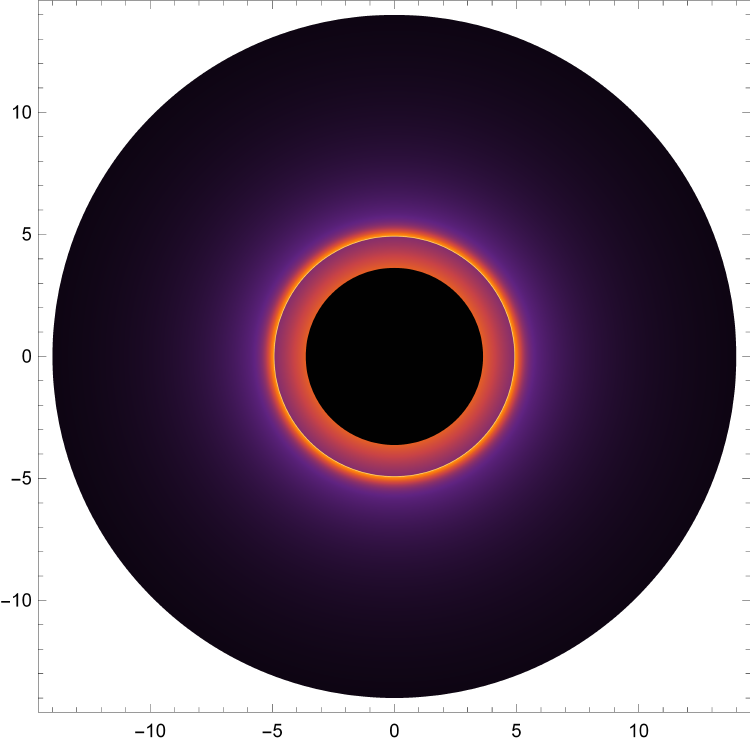}
		\end{minipage}
	}%
	\caption{The observed specific intensities and two-dimensional images of the BH in Horndeski theory for three emission models of geometrically thin accretion disks. We set $\gamma = -0.3$ and $M=1$.}
	\label{densityplot1}
\end{figure}

To enable a direct comparison with the BH images provided by the EHT, we apply a Gaussian filter with a width of $20 \mu$as to smooth the images presented in Figs.~\ref{Iobdensity4}, \ref{Iobdensity5}, and \ref{Iobdensity6}~\cite{EventHorizonTelescope:2019dse,Carballo-Rubio:2022aed,Psaltis:2020cte}, and then we generate Fig.~\ref{blur}. It can be observed that the application of the filter results in the removal of the distinct features associated with the lensing ring and the photon ring. Consequently, only a blurred image of the direct emission remains, with no change in its size. This suggests that, for the current resolution of the EHT, it is challenging to obtain more information about the lensing ring and photon ring in the BH images within the framework of Horndeski theory. 

\begin{figure}[!ht]
	\centering
	\subfigure[ \label{blur1}]{
		\begin{minipage}[t]{0.33\linewidth}
			\centering
			\includegraphics[width=4cm,height=4cm]{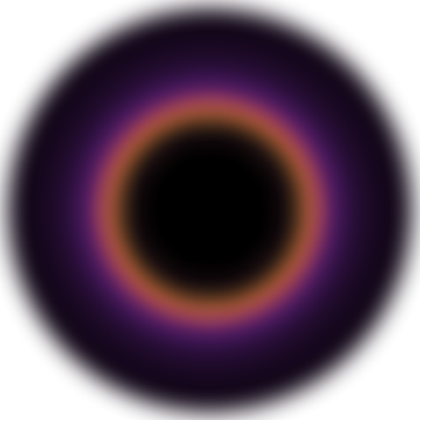}
		\end{minipage}%
	}%
	\subfigure[ \label{blur2}]{
		\begin{minipage}[t]{0.33\linewidth}
			\centering
			\includegraphics[width=4cm,height=4cm]{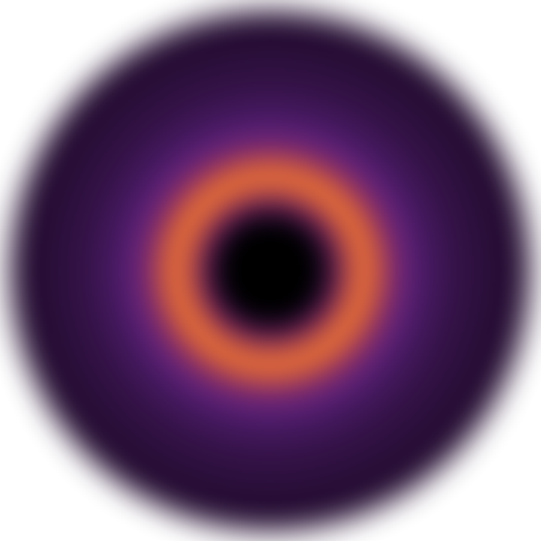}
		\end{minipage}%
	}
	\subfigure[ \label{blur3}]{
		\begin{minipage}[t]{0.33\linewidth}
			\centering
			\includegraphics[width=4cm,height=4cm]{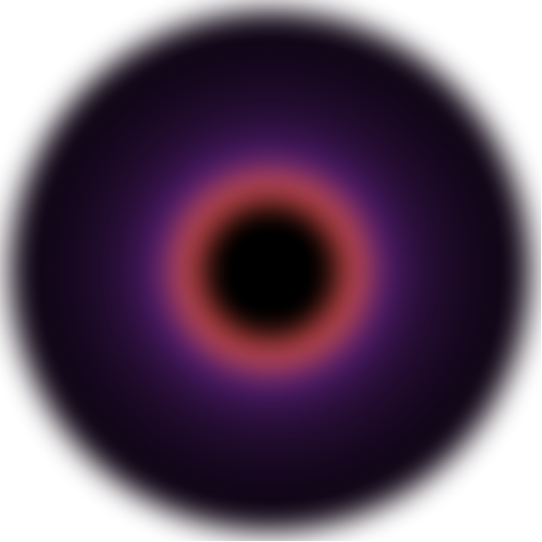}
		\end{minipage}%
	}
	\caption{The blurred two-dimensional images with a Gaussian filter with a width of $20 \mu$as.}
	\label{blur}
\end{figure}

\section{Conclusions}\label{sec:5}

In this paper, we investigate the observational features of the BH in Horndeski theory. First, we analyze the motion of photons in the BH spacetime background and derive the effective potential of photons, critical impact parameter, and photon sphere radius. We find that as the parameter $\gamma$ increases, the critical impact parameter, photon sphere radius, and event horizon radius also increase. We plot the trajectories of photons moving on the equatorial plane of the BH. Using the EHT observational data for $\mathrm{M} 87^*$ and $\mathrm{Sgr} \mathrm{A}^*$, we constrain the parameter $\gamma$ to $-0.46 \lesssim \gamma \lesssim 0.31$ at the $1\sigma$ confidence level. 

Next, we study the QNMs of a massless scalar perturbation field in the BH background by the 6th-order WKB method and the eikonal limit, respectively. The results of the two methods are consistent. Based on EHT observational data for $\mathrm{M} 87^*$ and $\mathrm{Sgr} \mathrm{A}^*$, we compute the QNMs for different values of the parameter $\gamma$. We find that as the parameter $\gamma$ increases, both the real part and negative imaginary part of the QNM frequencies decrease monotonically, which means that a smaller parameter $\gamma$ corresponds to a higher oscillation frequency of the scalar wave. However, as the parameter $\gamma$ increases, the shadow radius exhibits a monotonically increasing trend, which is opposite to the observed change in the real part of the QNM frequencies. We also calculate the frequency ranges of the fundamental modes of $\mathrm{M} 87^*$ and $\mathrm{Sgr} \mathrm{A}^*$ in Horndeski theory at the $1\sigma$ confidence level, which are constrained to $2.4\times 10^{-6}$ Hz $\lesssim f \lesssim 2.7\times 10^{-6}$ Hz and $3.6\times 10^{-3}$ Hz $\lesssim f \lesssim 4.1\times 10^{-3}$ Hz, respectively.

Finally, we study the optical appearances of the BH surrounded by a geometrically thin accretion disk. Based on the total number of photon orbits around the BH, we classify the light rays into three categories: direct emission, lensing ring, and photon ring. Then, the impact parameter can also be be divided into three categories according to  the classification of the light rays. We find that increasing values of the parameter $\gamma$ will cause the peak of the total number of photon orbits to shift towards the positive horizontal axis. Therefore, the sizes of the direct emission, lensing ring, and photon ring increase with the parameter $\gamma$. By analyzing the first three transfer functions, we find that the first transfer function corresponds to the direct emission with a demagnification factor close to 1, and so it contributes the majority of the total flux. The second transfer function corresponds to the lensing ring, with a larger demagnification factor, producing a reduced image of the back side of the disk and contributing a smaller portion of the total flux. The third transfer function corresponds to the photon ring with an extremely large demagnification factor, producing a highly reduced image of the front side of the disk and contributing minimally to the total flux. With the aforementioned analysis, we study the optical appearances of the BH under three emission models, i.e., the emission originating from the innermost stable circular orbit, event horizon radius, and photon sphere radius, respectively. In the first emission model, we find that the direct emission, lensing ring, and photon ring are distinct. In the second and third emission models, we observe the absence of a distinct photon ring, and the lensing ring overlaps with the direct emission. Thus, the optical appearances of the BH is heavily dependent on the choice of the emission model. Applying a Gaussian filter with a width of $20 \mu$as to smooth the optical appearances of the BH and comparing them with the images provided by the EHT, we find that the filter will remove the features of the lensing ring and photon ring, leaving only a blurred direct emission without any alteration in its size. This suggests that, given the current resolution of the EHT, it is challenging to obtain detailed information about the lensing ring and photon ring, and the actual image of the BH is largely influenced by the structure of the accretion disk. In future research, we aim to extend the analysis to rotating BHs, in order to impose more stringent constraints on Horndeski theory.

\vspace{10pt}

\noindent {\bf Acknowledgments}

\noindent
This work was supported by the National Key Research and Development Program of China (Grant No. 2021YFC2203004), the Natural Science Foundation of Chongqing (Grant No. CSTB2023NSCQ-MSX0103), and the National Natural Science Foundation of China (Grant No. 12347101).

\end{document}